\documentclass[12pt]{article}
\usepackage{amsmath,amssymb,graphicx,float,epsf}
\def\RR{\hbox{I\kern-.2em\hbox{R}}}
\newcommand{\qed}{\hbox to 0pt{}\hfill$\rlap{$\sqcap$}\sqcup$ \vspace{3mm}}
\numberwithin{equation}{section}
\restylefloat{figure}

\usepackage{graphicx} %%For loading graphic files
\usepackage{amsmath}
\usepackage{amsthm}
\usepackage{amsfonts}
\usepackage{enumerate}
\usepackage{enumitem}

\usepackage{graphicx}
\usepackage{tikz}
\usetikzlibrary{mindmap,backgrounds,arrows.meta}
\usetikzlibrary{shapes.geometric, arrows}
\tikzstyle{rect} = [draw, rectangle, fill=blue!20, text width=6em, text centered, minimum height=2em]
\tikzstyle{elli} = [draw, ellipse, fill=red!20, minimum height=2em]
\tikzstyle{circ} = [draw, circle, fill=white!20, minimum width=8pt, inner sep=5pt]
\tikzstyle{diam} = [draw, diamond, fill=white!20, text width=6em, text badly centered, inner sep=0pt]
\tikzstyle{line} = [draw, -latex']

\usepackage{a4wide} %%Smaller margins = more text per page.
\usepackage{authblk} % for multiple authors affiliation
\usepackage{varioref}
\usepackage{hyperref}
\hypersetup{colorlinks=true, linkcolor=violet, citecolor=red, urlcolor=blue} %olive,magenta,orange,purple,teal

\allowdisplaybreaks
\usepackage[percent]{overpic}
\usepackage{subcaption}
\usepackage[superscript,biblabel]{cite}

\begin{document}
% Don't want date printed
\date{}

\title{\bf  Vaccine efficacy and SARS-CoV-2 control in California and U.S. during the session 2020–2026: A modeling study}

\author[1]{\small Md. Shahriar Mahmud\thanks{Email: prism.shahriar@gmail.com}}
\author[2,3]{\small Md. Kamrujjaman \thanks{Corresponding author: M. Kamrujjaman, email: kamrujjaman@du.ac.bd, Tel.: +880-1-553-458-910, \\ 
ORCID ID: \url{https://orcid.org/0000-0002-4892-745X}}}
\author[4]{\small Md. Mashih Ibn Yasin Adan\thanks{Email: mdadan1081@gmail.com }}
\author[5]{\small Md. Alamgir Hossain\thanks{Email: alamgir.jnu10@gmail.com  }}
\author[6]{\small Md. Mizanur Rahman\thanks{Email:  mizanur.rahman@r.hit-u.ac.jp}}
\author[7]{\small Md. Shahidul Islam\thanks{Email: mshahidul11@yahoo.com }}
\author[8]{\small Muhammad Mohebujjaman\thanks{Email: m.mohebujjaman@tamiu.edu}}
\author[9,10]{\small Md. Mamun Molla \thanks{Email: mamun.molla@northsouth.edu}} 

%\author[5]{\small Shohel Ahmed\thanks{Email: shohel2@ualberta.ca}}
%\author[1,2]{\small Md Nazmul Hassan\thanks{Email: md.nazmul.hassan@ttu.edu}}
%\author[1]{\small Kaniz Fatema Nipa\thanks{Email: kaniz.fatema.nipa@ttu.edu}}

\affil[1]{\footnotesize Department of Computer Science and Engineering, State University of Bangladesh, Dhaka 1205, Bangladesh}
\affil[2,4,7]{\footnotesize Department of Mathematics, University of Dhaka, Dhaka 1000, Bangladesh}
\affil[3]{\footnotesize Department of Mathematics and Statistics, University of Calgary,  Calgary, AB, Canada}
%\affil[2]{\footnotesize Department of Mathematics and Statistics, University of Calgary, 2500 University Dr., Calgary, AB, Canada, T2N 1N4}
\affil[5]{\footnotesize Computational Biology Research Lab (CBRL), Department of Pharmacy, Jagannath University, Dhaka 1100, Bangladesh}
%\affil[5]{\footnotesize Mathematical and Statistical Sciences, University of Alberta,  Edmonton, AB, T6G 2R3, Canada}
\affil[6]{\footnotesize Hitotsubashi Institute for Advanced Study, 	Hitotsubashi University,  Naka Kunitachi Tokyo 186-8601, Japan}
\affil[8]{\footnotesize Department of Mathematics and Physics, Texas A\&M International University, Laredo, TX 78041, USA}
\affil[9]{\footnotesize Department of Mathematics \& Physics, North South University, Dhaka 1229, Bangladesh} 
\affil[10]{\footnotesize Center for Applied Scientific Computing (CASC), North South University, Dhaka 1229, Bangladesh}

\maketitle

\vspace{-0.5cm}
%\noindent\rule{6.35in}{0.02in}\\
\noindent{\bf \large Abstract}\\%Summary}\\	
\textbf{Background:}  Besides maintaining health precautions, vaccination has been the only prevention from SARS-CoV-2, though no clinically proved 100\% effective vaccine has been developed till date. At this stage, to withhold the debris of this pandemic- experts need to know the impact of the vaccine efficacy rate's threshold and how long this pandemic may extent with vaccines that have different efficacy rates. In this article, a mathematical model study has been done on the importance of vaccination and vaccine efficiency rate during an ongoing pandemic.\\ [1.5mm]
\textbf{Methods:}  We simulated a five compartment mathematical model to analyze the pandemic scenario in both California, and whole U.S.. We considered four vaccines, Pfizer (95\%), Moderna (94\%), AstraZeneca (79\%), and Johnson \& Johnson (72\%), which are being used rigorously to control the COVID-19 pandemic, in addition with two special cases: a vaccine with 100\% efficacy rate and no vaccine under use. We considered four vaccines, Pfizer (95\%), Moderna (94\%), AstraZeneca (79\%), and Johnson \& Johnson (72\%), which are being used 
rigorously to control the COVID-19 pandemic. COVID-19 related data of California, and U.S. were used in this study. \\ [1.5mm]
\textbf{Findings:} Both the infection and death rates are very high in California. Our model suggests that the pandemic situation in California will be under control in the last quartile of the year 2023 if frequent vaccination is continued with the Pfizer vaccine. During this time, six waves will happen from the beginning of the immunization where the case fatality and recovery rates will be 1.697\% and 98.30\%, respectively. However, according to the considered
model, this period might be extended to the mid of 2024 when vaccines with lower efficacy rates are used. On the other hand, the daily cases and deaths in the U.S. will be contained at the end of 2026 with multiple waves. Although the number of susceptible people will reach zero in the mid of 2028, there is less chance to stop the vaccination program. Pfizer vaccine recipients will get down to zero at the end of 2024 as the SARS-CoV-2 situation will be under control, while a significant number of people will be vaccinated with Pfizer, Moderna, and AstraZeneca vaccines till the mid of 2028. According to this study, the infected and exposed cases will be under control at the end of 2027 and the mid of 2028, respectively. Remarkably, we considered that the pandemic situation is under control when the number of active cases is below 100, and this situation remains for more than 20 weeks. \\ [1.5mm]
\textbf{Interpretation:}  The more effective a vaccine, the less people suffer from this malign infection. Although specific groups of people get prioritized initially, mass vaccination is needed to control the spread of the disease.  
\\[1.5mm]
%\\[-3mm]
%\textbf{Funding:} The research by M. Kamrujjaman was partially supported by a TWAS grant 2019\_19-169 RG/MATHS/AS\_I. \\ [-3mm]

\noindent{\it \footnotesize Keywords}: {\small Vaccine;  Model;  California;  Control measurement; 
	SARS-CoV-2. }\\
\noindent{\it \footnotesize AMS Subject Classification 2010:} 92D25; 92D30; 97M60; 97M99.\\
\noindent
%\noindent\rule{6.35in}{0.02in}

%\section*{Research in context/Highlights}
%\begin{enumerate}
%	\item Since vaccination is the only clinical way to prevent SARS-CoV-2 virus, the impact of vaccine worths a look. 
%	\item This article studies the results of the effectiveness of some widely used vaccines through a five compartment mathematical model. \item The model forecasts the SARS-CoV-2 scenarios for both the California state and the U.S. 
%	\item The findings of this study are very trivial and show that a vaccine with or more than 95\% effectiveness would make a constructive change in the control. 
%	\item A perfect vaccine (100\% effective) and mass vaccination program is mandatory to withhold the crucial situation worldwide.
%\end{enumerate}

\clearpage

\section*{Introduction}
Severe Acute Respiratory Syndrome Coronavirus 2 (SARS-CoV-2) induced disease COVID-19 has a devastating impact on every sector of human life \cite{Ref_2}. A Global emergency was declared due to this COVID-19 outbreak on January 30, 2020, by the World Health Organization (WHO) \cite{Ref_3}. To contain the spread of the virus \cite{Ref_4}, governments initially followed different approaches, including the shutdown of the borders, restrictions on both local and international travel, quarantine, and nationwide lockdown \cite{Ref_5, Ref_6}.  The education system from pre-school to university education has been affected by COVID-19, and in most cases, authorities have closed the educational institutions. According to UNESCO, approximately 900 million students have faced the detrimental effect of this closure \cite{Ref_7}. An international survey has been carried out by the International Association of Universities (IAU) to estimate the impact of this outbreak on the global education system. 78\% of participants are convinced by the fact that COVID-19 will engender an adverse influence on the number of students who will enroll in the upcoming academic year. Approximately half of the respondents (46\%) think both local and international students will face problems due to this ongoing pandemic \cite{Ref_8}. As well as the influence on undergraduate education \cite{Ref_9}, the most critical impact on postgraduate research is that many irrelevant COVID research topics have been halted. The national funding body for health research in the U.K. stopped all non-COVID research to facilitate the opportunity for clinically trained staff to return to the frontline \cite{Ref_10}. In the USA, similar action was taken by the National Institute for Health to stop all non-critical research. Apart from healthcare research, many institutions of Harvard University closed laboratories in the Faculty of Arts and Sciences in order to abstain from researching humanities and social sciences \cite{Ref_11, Ref_12}.
Moreover, the concern is raised about the canceled scientific conferences. Conferences are now taking place on online platforms, and this medium is not that viable for networking compared to physical conferences as these conferences are considered significant to many fields of scientific research and opportunities for collaboration \cite{Ref_13}.  The approach to social isolation due to COVID-19 has had a severe outcome on the psychological and mental health of the people in the society. Suicide, self-harm, substance misuse, domestic and child abuse are many of the presumed repercussions of this isolation. Since March 9, 2020, in the U.K., 4000 offenders have been arrested due to domestic abuse that is equating to 100 per day, indicating the negative result of social isolation \cite{Ref_14, Ref_15}. According to a report \cite{Ref_16}, France and the USA have seen 32-36\% and 21-35\% surge in domestic abuse. In the U.K., there has been an increase in domestic abuse hotline calls by 25\% and 75\% enhancement in the Google search regarding support for domestic abuse. This abuse is mainly due to the reduction of the opportunities for support, enhanced contact to manipulative relationships, and disaster-related indoor uncertainty \cite{Ref_15, Ref_16}.

However, the outbreak of COVID-19 by coronavirus in Wuhan is not the first outbreak by the virus. Before this, there were two outbreaks known as Severe Acute Respiratory Syndrome (SARS) in China in 2002, followed by Middle East Respiratory Syndrome (MERS) in 2012 in Saudi Arabia \cite{Ref_17, Ref_18}. To treat the 2003 SARS infection, scientists quickly developed different vaccines, and VRC-SRSDNA015-00VP was one of them, which is a DNA vaccine that consists of a circular plasmid DNA macromolecule (VRC-8318) \cite{Ref_19}. Phase I human trial started after 17 months of evaluating the preclinical safety and efficacy of the vaccine candidate \cite{Ref_20}. Ten healthy volunteers aged 21-49 took part in this trial from December 13, 2004, to May 2, 2005. Four milligrams (4 mg) of the vaccine were injected three times at four weeks intervals, and the subjects were monitored for 32 weeks \cite{Ref_21, Ref_22, Ref_23}. However, 9 participants out of 10 completed the recommended doses and, one of them was withdrawn from the list to treat poison ivy contact dermatitis after receiving the second vaccination \cite{Ref_24}. While analyzing the vaccine's effectiveness, it was found that the VRC DNA SARS vaccine generated response against T-cell and antibody of SARS virus and neutralized antibody in 8 out of 10 subjects. After completing the phase I clinical trial of the VRC DNA SARS vaccine, it was declared that the safety and tolerability were within the favorable range \cite{Ref_21, Ref_22, Ref_23, Ref_25, Ref_26, Ref_27, Ref_28}. Unfortunately, no data/information of this vaccine regarding the further trial phases is available. For MERS coronavirus, GLS-5300 was the first vaccine that entered the phase I clinical trial, and it was also a DNA vaccine. Initially, the vaccine was given to 75 adults, from 18 to 50 years old. 0.67 mg, 2 mg, or 6 mg of GLS-5300 vaccine was administered on the 0, 28th, and 84th days. Safety assessment of the volunteers of this study was monitored up to 48 weeks after completing the doses \cite{Ref_29}. After analysis of phase I clinical trial results, it was found that 93\% of participants experienced a reaction at the injection site and 92\% experienced pain. Infection was the most common unwanted adverse effect reported in 36\% of participants. Also, the antibody neutralizing effect was observed in 27 volunteers at 14th weeks, 25 volunteers at 24th weeks, and 2 volunteers at 60th weeks \cite{Ref_30}.  

To date, there are many outbreaks of infectious diseases apart from these outbreaks of infectious disease by a coronavirus. Cholera is such a kind of disease that is induced by the bacterium \textit{Vibrio Cholerae} and the intestine is usually damaged when infected by this bacterium. Although this disease is endemic in few Asian and African countries, several cholera epidemics have been noticed worldwide. It is estimated that about 1.3-4 million people get affected by this disease, and among them, 21,000-143,000 face death per year \cite{Ref_31}. In 2017, 1,227,391 cases and 5,654 deaths were reported in 34 countries worldwide \cite{Ref_32}. Rapid dehydration and the inconsistency of electrolytes in our bodies are common scenarios of this disease. Without swift action, cholera might be the reason for the death of a person due to dehydration within a few hours \cite{Ref_33}. Enhanced water and sanitation systems are indeed useful to prevent the disease, but the most effective way is a vaccination. This strategy is following to contain the spread of this infectious disease, particularly in Haiti \cite{Ref_34}. Among 200 serotypes of the bacterium, two of them are responsible for the disease cholera, and they are \textit{V. cholerae} O1 and O139. Dukoral (WC-rBS) and Shanchol are two available vaccines for cholera that are administered orally.  mORCVAX is an identical vaccine similar to Shanchol \cite{Ref_35}. Dukoral is available in more than sixty countries worldwide, and it consists of the deadly strains of \textit{V. cholerae} and recombinant cholera toxin beta (CTB) \cite{Ref_36, Ref_37}. Whereas Shanchol contains three deadly strains of \textit{V. cholerae} O1 and single O139, but it does not contain any CTB like Dukoral \cite{Ref_38}. Different doses of Dukoral are given depending on age. Those who are more than 6 years old are given 2 doses, whereas children below 6 years receive three doses of this vaccine at least a week interval \cite{Ref_37}. On the other hand, two doses of Shanchol are administered to those who are $ \ge $ 1-year-old at 14 days apart \cite{Ref_39}. These vaccines' most potent protective action is observed against the disease in the first 2 years of vaccination. According to the efficacy report of these vaccines, their efficacy ranging from 86\% to 66\% at 4-6 months, 62\% to 45\% at 1 year and, 77\% to 58\% at 2 years \cite{Ref_35}. Another infectious disease is malaria which roughly abounds in 91 countries globally. Out of 120 species, only six species of Plasmodium are responsible for infecting the human \cite{Ref_40}. According to the latest World malaria report published by WHO on November 30, 2020, there were 229 million malaria cases and 409,000 death worldwide. Although the number of deaths was comparatively lower compared to 2018, which was 411,000. African countries were responsible for almost 94\% of all malaria cases, and deaths \cite{Ref_41}. To date, only RTS, S/AS01 is the most effective anti-malarial vaccine studied the most. Interestingly, this vaccine is the result of a collaboration started in the 1980s between the Walter Reed Army Institute of Research in the USA and GSK biologicals \cite{Ref_42}. An experiment of this vaccine was conducted initially in African children, and the significant effect was measured against infection of 
\textit{P. falciparum} and this species is regarded as the deadliest parasite worldwide and prevalent in Africa. The vaccine was capable of resisting the malarial infection in 4 out of 10 cases over 4 years in children who received four doses during the clinical trial period \cite{Ref_41}. 

To eradicate any pandemic disease, vaccination is the most prominent way. The immune system is boosted by vaccination. Neither any diseases nor vaccination indeed provides immunity. Vaccination becomes successful because, in most cases, every disease has a recovered/immune stage \cite{Ref_43}. Vaccination plays an integral role in ameliorating of living and health standards of people \cite{Ref_44, Ref_45}.  After the emergence of Covid-19, scientists started developing vaccines relentlessly worldwide. Finally, for the first time on November 9, 2020, NEW YORK \& MAINZ, GERMANY (BUSINESS WIRE) Pfizer Inc. (NYSE: PFE) and BioNTech S.E. (Nasdaq: BNTX) announced their vaccine BNT162b2 that has made based on the mRNA as effective against COVID-19. This result was from the phase III clinical trial of the vaccine where the volunteers did not have any prior symptoms of SARS-CoV-2 infection.
Forty-three thousand five hundred thirty-eight volunteers were recruited in this part of the trial, and it was found that the vaccine was 90\% effective in resisting the viral infection. Also, the analysis confirmed 94 COVID-19 cases in participants during this trial period \cite{Ref_46}. Later on December 10, 2020, an article published in The NEW ENGLAND JOURNAL of MEDICINE regarding the safety and efficacy of the vaccine of further study where; it was found that the vaccine is 95\% capable of protecting the disease. In this part of the trial, 43,548 took part; among them, 43,448 were injected. 21,720 participants were given either BNT162b2, and 21,728 received placebo. All of them received two doses, and the second dose was given after 21 days of the first dose, and the concentration was 30 $ \mu $g.  After the first dose, 10 severe COVID-19 cases were raised, and 9 related to those assigned to placebo, and the remaining one was BNT162b2 recipient.  After 7 days of administration of the second dose, there were 8 confirmed cases of COVID-19 among the participants who received BNT162b2 whereas 162 cases were observed in the case of placebo. The most common side effects were pain at the injection site, fatigue, headache, muscle pain, chills, joint pain, and fever which remained a few days. Notably, most volunteers had these side effects after completing the final dose \cite{Ref_47}.  More than a month later after the first declaration, on December 11, 2020, the U.S. Food and Drug Administration (FDA) approved the vaccine for emergency use authorization (EAU) in individuals who are 16 years old and older to prevent the disease COVID-19 \cite{Ref_48}. The USA, the UK, Canada, Bahrain, Mexico, and Singapore are some countries that approved this vaccine rapidly to use throughout the country after getting approval from the U.S. FDA \cite{Ref_49, Ref_50}. Pfizer has announced that they are expecting to produce 50 million doses at the end of this year and 1.3 billion by the end of the following year \cite{Ref_51}.

Another vaccine that is being used extensively is ChAdOx1 nCoV-19 (AZD1222) that Oxford-AstraZeneca has developed. Report from the clinical trial has shown that this vaccine can also protect against the SARS-CoV-2 infection. From April 23 to November 4, 2020, 23,848 volunteers took part in a clinical trial, and 11636 (7548 in the U.K., 4088 in Brazil) were taken randomly to analyze the interim primary efficacy. Participants received either ChAdOx1 nCoV-19 vaccine or control (meningococcal group A, C, W, and Y conjugated vaccine or saline). Two doses of the vaccine were injected composed of $ 5\times 10^{10} $ units of viral (Standard dose: SD/SD cohort). 70.4\% found the overall vaccine efficacy. Twenty-one days after the first administration of the first dose, 10 cases of COVID-19 induced who received the control where two were regarded as severe cases, including one death. One hundred seventy-five severe adverse reactions were observed in 168 people, and among them, 84 were associated with ChAdOx1 nCoV-19, and 91 in control \cite{Ref_52}. 
The latest report on the efficacy published on March 25, 2021, suggests that the Oxford-AstraZeneca vaccine efficacy is $76\%$. The interim analysis of the vaccine efficacy showed that it was $79\%$ effective against the virus. This result was from a trial where 32,449 adults participated from the U.S., Peru, and Chile. During the trial period, no one was reported to get admitted to the hospital or died, although $60\%$ of the participants had previous health issues like diabetes or obesity \cite{ref_52, ref_53}. However, recently a severe issue, rare-blood clotting, has been reported due to this vaccine. This happens mostly in women age over 55 years old. Considering this critical issue, 20 European countries have halted vaccination \cite{ref_53}. Among these countries, Denmark has stopped using the vaccine completely. Due to this, 2.4 million doses of vaccines will be withdrawn until the following announcement. The Danish authority decided to do this after reporting two cases of blood-clotting, one of them was fatal in a 60 years old woman \cite{ref_54}.
On the other hand, Germany has suspended the regular use of the vaccine but decided to continue using it for people over 60.  This decision was made when few cases of rare blood clotting were observed for younger people \cite{ref_55}. Sweden, Latvia, Italy, Spain, France, Luxembourg, Cyprus, Portugal, Slovenia, Netherlands, Ireland are some other European countries that have suspended the use of the Oxford-AstraZeneca vaccine \cite{ref_56}. 

Moderna COVID-19 vaccine mRNA-1273, developed by Moderna TX, Inc got approval for emergency use for the first time by the FDA on December 18, 2020 \cite{ref_57}. In phase, I clinical trial, 45 adults aged from 18 to 55 participated, and they were given two doses of the vaccine with a 28 days interval. They were divided into three different groups and administered 25 $\mu$g. 100 $\mu$g, or 250 $\mu$g per dose for each group. After the first dose of the vaccine, it was found that the higher doses were capable of inducing a higher effect (Geometric mean titer (GMT) was 40,227 in 25 $\mu$g, 109,209 $\mu$g, and 213,526 in 250 $\mu$g of the vaccine recipients, respectively). After the 2nd dose of the vaccine, a similar pattern was observed: a higher amount of the vaccine generated higher responses. At that time, the GMT were 2999,751, 782,719, and 1,192,154, respectively \cite{ref_58}. On the other hand, 30,420 volunteers from the U.S. participated in phase III clinical trial of mRNA-1273. They were divided into two groups (15,210 in each group) and given either the vaccine (100 $\mu$g) or placebo. Symptomatic illness was observed due to Covid-19 infection was detected in 185 and 11 volunteers who were given the placebo and vaccine, respectively \cite{ref_59}. The most common observed side effects of the vaccine were pain at the injection site $(92\%)$, fatigue $(70\%)$, headache $(64.7\%)$, muscle pain $(61.5\%)$, chills $(45.4\%)$, joint pain $(46.4\%)$, fever $(15.5\%)$, swelling at the injection site $(14.7\%)$, erythema at the injection site $(10\%)$, nausea, vomiting $(23\%)$, and lymphadenopathy, axillary tenderness $(19.8\%)$. No serious events were observed for the vaccine or placebo \cite{ref_57}. However, after the phase III clinical trial, the efficacy of the mRNA-1273 was found $94.1\%$ to prevent the severe illness due to the Covid-19 infection \cite{ref_59}. 

Ad26.COV2.S is another vaccine that the FDA has approved, U.S. developed by the Janssen Pharmaceutical Companies of Johnson and Johnson. The mechanism of action of this vaccine is different from the Pfizer and Moderna vaccine. According to the developer, it can combat moderate to severe Covid-19 infection in people age 18 and over.  Here, the modified DNA of adenovirus is used that can make a similar viral particle of SARS-CoV-2. When the vaccine is administered, the body will make an immune response against that particle. Eventually, this immune response will protect from the Covid-19 infection. Notably, the adenovirus is responsible for respiratory infection. From the phase III clinical trial, different efficacy was found. A single dose of the vaccine was $66\%$ and $67\%$ effective against moderate to severe-critical Covid-19 condition after 14 and 28 days of the administration, respectively.
Interestingly, higher efficacy was observed for the severe-critical condition that $77\%$ and $85\%$ after 14 and 28 days of vaccination, respectively \cite{ref_60, ref_61}. However, a report published on April 13, 2021, reveals that the U.S. has stopped using this vaccine due to forming blood clots. Six cases were found having blood clotting combined with low platelets. All of them were females aged from 18 to 48 years old. This condition appeared after 6 to 13 days of vaccination \cite{ref_62}. 

Apart from these four vaccines, few other vaccines got approval and using in many countries. The list of those vaccines is given below [Updated on April 20, 2021] \cite{ref_63}:

\begin{table} [H]
	%	\caption{List of vaccines that have been approved for emergency use}
	\begin{tabular}{|l|l|l|l|l|}
		\hline
		{\bf Developer}	&{\bf Vaccine name}	&{\bf Type} &{\bf Efficacy} &{\bf Doses} \\ \hline
		Gamaleya (Russia)	&Sputnik V	&Adenovirus  &$91.6\%$ &\vtop{\hbox{\strut 2 doses, 3 weeks}\hbox{\strut  apart}}%2 doses, 3 weeks apart 
		\\\hline
		\vtop{\hbox{\strut CanSino Biologics}\hbox{\strut (China)}} 	&Convidecia	&Adenovirus  &$65.28\%$ & Single dose 
		\\\hline
		\vtop{\hbox{\strut Vector Institute}\hbox{\strut (Russia)}}  	&EpiVacCorona 	&Peptides  &Unknown &\vtop{\hbox{\strut 2 doses, 3 weeks}\hbox{\strut  apart}}%2 doses, 3 weeks apart 
		\\\hline
		Sinopharm (China)	&BBIBP-CorV	 &Inactivated   &$79.34\%$ &\vtop{\hbox{\strut 2 doses, 3 weeks}\hbox{\strut  apart}}%2 doses, 3 weeks apart 
		\\\hline
		\vtop{\hbox{\strut Sinovac Biotech}\hbox{\strut  (China)}} 	&CoronaVac 	&Inactivated   &\vtop{\hbox{\strut $50.65\%$ in Brazil,}\hbox{\strut $91.25\%$ in Turkey}}%$50.65\%$ in Brazil,   $91.25\%$ in Turkey 
		&\vtop{\hbox{\strut 2 doses, 2 weeks}\hbox{\strut  apart}}%2 doses, 3 weeks apart 
		\\\hline
		\vtop{\hbox{\strut Bharat Biotech}\hbox{\strut  (India)}} 	&Covaxin	&Inactivated   &$80.6\%$ &\vtop{\hbox{\strut 2 doses, 4 weeks}\hbox{\strut  apart}}%2 doses, 3 weeks apart 
		\\\hline
	\end{tabular}
	\caption{List of vaccines that have been approved for emergency use.}
\end{table}

Scientists are investigating 89 vaccines where 52 are in the phase I clinical trial and 37 are in phase II clinical trial stage. On the other hand, 23 vaccines have reached the phase III stage \cite{ref_63}.

\section*{Model overview}
The mathematical model in infectious disease is one of the most crucial issues in epidemiology. From mathematical models, we get translucent ideas about the pattern of disease behavior.

\subsection*{Model formulation and parameter description}
In this study, we propose a non-demographic  SIR type vaccination model (Susceptible-Infectious-Recovered) for the dynamics of SARS-CoV-2 infectious disease.

We assume that among any locality, almost all active populations are susceptible at the beginning of the virus circulation since a sudden introduction of an unknown and highly contagious virus takes few days to make the population conscious of its consequences. Meanwhile, specific types of contacts between susceptible and already infected and infectious individuals keep spreading the infection silently. When the virus-induced disease has started showing symptoms and mass fatality begins as a consequence, it is almost too late. It becomes impossible to identify and isolate all infected and infectious individuals from the susceptible population.   

After close contact with a SARS-CoV-2 patient, a susceptible individual may have two different scenarios: His enough precautions and/or strong immune system may immediately refuse to let the sufficient amount of virus enter in his body, and so that body is still not infected rather susceptible yet. Another scenario may happen; he got to take many viruses in his body because of his ill knowledge and/or inadequate safety measures, and his immune system starts fighting against the virus. This person must appear for a clinical test. A positive clinical test result is the confirmation of a SARS-CoV-2 infection case. For this kind of infection, a clinical test supports disease diagnosis, not treatment. If there are not enough facilities to provide an immediate test to the susceptibles who have come to contact with the confirmed infected population, or who have been asymptomatic throughout their entire infection period and who have been pauci-symptomatic (sub-clinical), pre-symptomatic (going to develop symptoms later), or post-infection (with still detectable viral RNA fragments from an earlier infection) \cite{bmj}. Muge Cevik {\em et. al.} showed that SARS-CoV-2 infected individuals might become infectious one to two days before the exposure of symptoms and also can continue to be infectious up to the seventh day after that \cite{Cevik1}. Now, it has been proved that symptomatic and pre-symptomatic transmission have a major role in the proliferation of SARS-CoV-2 than the truly asymptomatic transmission \cite{buitrago,byam,Qui,Cevik2}.

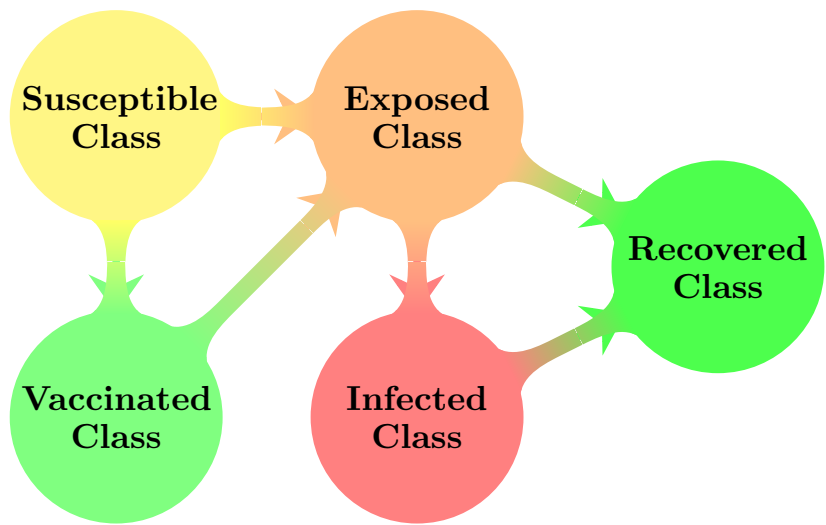
\begin{figure} [H]
	\centering
	\begin{tikzpicture}[small mindmap, outer sep=0pt, text=black]
		
		\begin{scope}[concept color=yellow!100!white!60]
			\node (s) at (0,0) [concept, scale=1.2] {\bf \small Susceptible Class}
			[counterclockwise from=210];
		\end{scope}
		
		\begin{scope}[concept color=orange!50]
			\node (e) at (4,0) [concept, scale=1.2] {\bf \small Exposed Class}
			[clockwise from=70];
		\end{scope}
		
		\begin{scope}[concept color=green!50]
			\node (v) at (0,-4) [concept, scale=1.2] {\bf \small Vaccinated Class}
			[clockwise from=70];
		\end{scope}
		
		\begin{scope}[concept color=red!50]
			\node (i) at (4,-4) [concept, scale=1.2] {\bf \small Infected Class}
			[clockwise from=70];
		\end{scope}
		
		\begin{scope}[concept color=green!70]
			\node (r) at (8,-2) [concept, scale=1.2] {\bf \small Recovered Class}
			[clockwise from=70];
		\end{scope}
		
		%	    \begin{scope}[concept color=green!40!black!30]
		%	    	\node (s) at (-2,0) [concept, scale=1.2] {\bf \small Susceptible Class}
		%	    	[counterclockwise from=210]
		%	    	%			child { node[concept] {L.1}  }
		%	    	%			child { node[concept] {L.2} }
		%	    	;
		%	    \end{scope}
		
		\path (s) to[circle connection bar switch color=from (yellow!100!white!60) to (orange!50)] (e) ;
		\path (s) to[circle connection bar switch color=from (yellow!100!white!60) to (green!50)] (v) ;
		\path (v) to[circle connection bar switch color=from (green!50) to (orange!50)] (e) ;
		\path (e) to[circle connection bar switch color=from (orange!50) to (red!50)] (i) ;
		\path (e) to[circle connection bar switch color=from (orange!50) to (green!70)] (r) ;
		\path (i) to[circle connection bar switch color=from (red!50) to (green!70)] (r) ;
		
		\begin{pgfonlayer}{background}    
			\draw 
			[concept connection,->,orange!50,shorten >= -0.15pt,-{Stealth[angle=70:1pt 6]}] 
			(s) to (e);    
			\draw 
			[concept connection,->,green!50,shorten >= -0.15pt,-{Stealth[angle=70:1pt 6]}] 
			(s) to (v);    
			\draw 
			[concept connection,->,orange!50,shorten >= -0.15pt,-{Stealth[angle=70:1pt 6]}] 
			(v) to (e);    
			\draw 
			[concept connection,->,red!50,shorten >= -0.15pt,-{Stealth[angle=70:1pt 6]}] 
			(e) to (i);    
			\draw 
			[concept connection,->,green!70,shorten >= -0.15pt,-{Stealth[angle=70:1pt 6]}] 
			(i) to (r);    
			\draw 
			[concept connection,->,green!70,shorten >= -0.15pt,-{Stealth[angle=70:1pt 6]}] 
			(e) to (r);
		\end{pgfonlayer}
		
	\end{tikzpicture}
	\caption{Compartmental model scheme visualization.}
	\label{model_scheme}
\end{figure}

Multiple pieces of research have found that the transmission rates can be 3 to 25 times higher for people with symptoms than those who are asymptomatic \cite{buitrago,Qui,Madewell,Koh}. A study in Wuhan casting almost 10 million people concluded with no proof of asymptomatic transmission till their study had published \cite{Cao}. 
Usually, the viral particles may shed from infectious people via talking, breathing, and/or coughing, where coughing may cause more viral particles to be spread than the other two mediums; which makes people with symptoms more contagious, as coughing is a major symptom of SARS-CoV-2 \cite{Chen}. 
In addition, pre-symptomatic and asymptomatic individuals naturally have more contacts than the isolated symptomatic individuals \cite{bmj}. 
Now it is clear that there are always some infected and so infectious people out there who are not intended to undergo the clinical test at the very primary stage and/or will keep spreading the infection as long as s/he gets his/her clinical result and has been taken to isolation/hospital.

%https://www.uchealth.org/today/the-truth-about-asymptomatic-spread-of-covid-19/

%https://www.bmj.com/content/bmj/371/bmj.m4851.full.pdf

In most of the cases ($97\%$ approximately worldwide \cite{wm}) for SARS- CoV-2 out-break, the symptomatic clinically tested confirmed infected individuals get recovered with proper health maintenance and regular medications. These recovered/discharged populations can hold herd immunity against the virus in maximum cases, while some recovered/discharged populations may be reinfected according to their immune ability and/or age. World Health Organization declared the reinfection case percentage may lie up to $ 0.01\% $ \cite{who_qa}. Moreover, till now, the SARS-CoV-2 virus has taken $ 3\% $ \cite{wm} lives of the clinically tested confirmed cases. Mainly, the old aged and people with the vulnerable immune system because of other illnesses, and people with unhealthy lifestyles are the major victims of not survival.   

In our model, we introduce a vaccinated compartment ($ V $) being activated from the day when an authorized, hence clinically effective vaccination program has been started in any specific locality or among any particular population/group. Since no vaccine may work $ 100\% $ effectively, so there remains a slight risk of being infected through a total dose of vaccination has been taken. We included this concept by considering the vaccine inefficacy parameter in the model. While in close contact with another infectious individual without proper health precautionary measures, this parameter leads a specific population to the infectious class/presymptomatic situation from the vaccinated compartment and then to the infection class. The primary focus of this study is to realize the role of vaccination in controlling the pandemic situation along with the importance of vaccine efficiency rate of vaccination in this aspect.   

%Recent scenario demonstrates that recovery individuals are wee grow up due to effective vaccination albeit few reports are depicted that some individuals are affected by the COVID-19 disease after taking vaccine. We take data from some well-known institution and online sources to depict the disease dynamics. Moreover, we introduce several parameters which may help to reduce the disease. From these values, we get clear conception about the patterns of disease behavior.
	
\section*{Numerical results}
Numerical simulation is the most incredible way to observe and present the outcomes of mathematical models. In this section, we will demonstrate the numerical results of our proposed model. Model prediction and actual data of infected cases of the SARS-CoV-2 virus give a better conception of this disease pattern. Moreover, we get a clear idea about forecasting and controlling the disease with the vaccine.

We use the Crank–Nicolson finite difference method for the model, and the graphical presentations are executed with MATLAB.

\subsection*{Case: California, U.S.}
California is one of the most important states of the United States. It is the third-largest state by size (423,970 $ km^2 $) with a total population of 39,512,223 (approximately on July 01, 2019) which makes it the most significant state by population \cite{census}. Corresponds to this data, California has also been affected by SARS-CoV-2 very hardly till the begging. On January 25, 2020, California reported its first SARS-CoV-2 case, where the first confirmed case in the U.S. was reported on January 21, 2020, by the case of a Washington resident who returned from Wuhan, China on January 15, 2020 \cite{ajmc,wiki_cal}. Now, California is on the top of the list among U.S. states with the highest $ 11.38\% $ confirmed cases and $ 10.65\% $ reported deaths of the total U.S. data till date \cite{google}. As the well-tuned data of our sample population is publicly available online, we collect the daily case data and fit our model with possible parameter values and observe the results in this subsection.

Moreover, California is offering the most advanced treatment for the SARS-CoV-2 virus. It has already started a vaccination program from December 15, 2020 \cite{Ref_61}. Nevertheless, it is very crucial to reach the vaccine to such a vast population overnight. So, the public health administration has taken some effective initiatives to provide vaccines to the people who need them the best. In the beginning, the authority decided that the people who have lived more than 50 years of their lives were being the most vulnerable against the virus. The front-line health workers are also straightforward to get infected, and some people have a weak immune system for other health issues. Californian health authority started assuring vaccines for these individuals in the first place. However, over time it also has declared that any 16+ aged individual will be eligible for vaccination from April 15, 2021  \cite{Ref_62}. Overall, the vaccination program is prioritized according to risk and age. Only high-risk patients and old-aged individuals are getting prioritized for vaccination, and younger people are comparatively at low risk to be infected by the COVID-19 virus. We consider the 30-day average vaccination number and imply for the numerical study.

The vaccination program is ongoing in hospitals, community vaccination sites, Doctor's offices, clinics, and pharmacies. Effective vaccine curbs against the SARS-CoV-2 virus spreading. It takes a few weeks after vaccination for the body to build protection against the virus. However, after taking the vaccine, it is still possible to get infected by the SARS-CoV-2 virus. After taking the full dose of the vaccine, it may work against the severe SARS-CoV-19 virus. But, the everyday new stain of the SARS-CoV-2 virus is being identified all around the world. The FDA, U.S. has demonstrated the safety and effectiveness of some vaccines by clinical trials and then authorized for public usage \cite{Ref_63, Ref_64}. Moreover, the authorized vaccines are up to $95\%$ (Pfizer) effective with a base of $ 72\% $ (Johnson \& Johnson) effectiveness against the SARS-CoV-2 disease. However, eventually, California is using the Pfizer and Moderna vaccine now. In this study, for different vaccine efficacy (`Pfizer' $95\%$, `Moderna' $94\%$, `AstraZeneca' $79\%$, and `Johnson $\&$ Johnson' $72\%$  \cite{Ref_57}), we illustrate the proposed model prediction in below figures which agree with the actual data. The United States vaccine safety system ensures that all vaccines are safe, though not entirely practical. The federal government is working on it \cite{Ref_65}.

\begin{figure} [H]
	\centering
	\includegraphics[width=0.48 \linewidth]{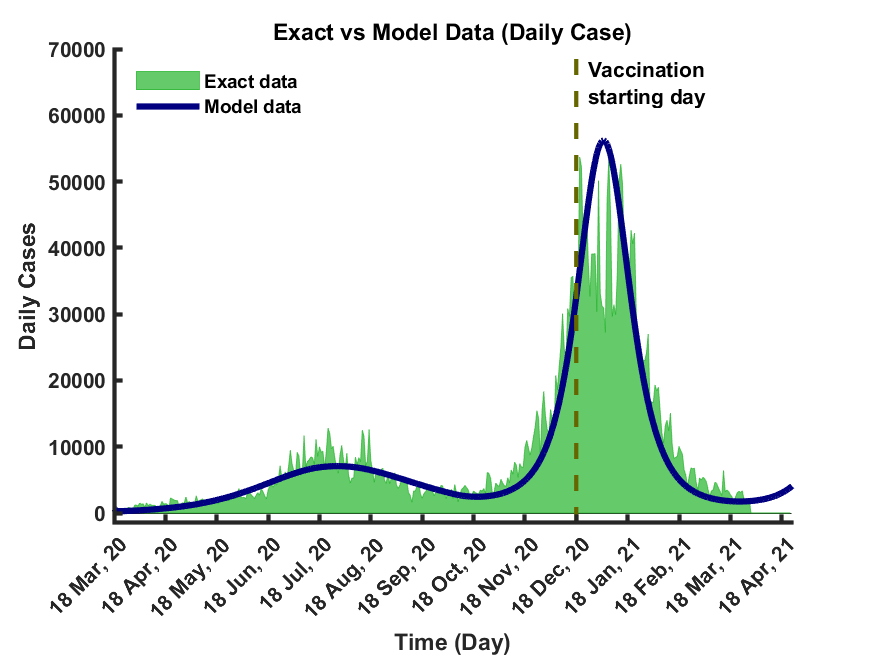}
	\includegraphics[width=0.48 \linewidth]{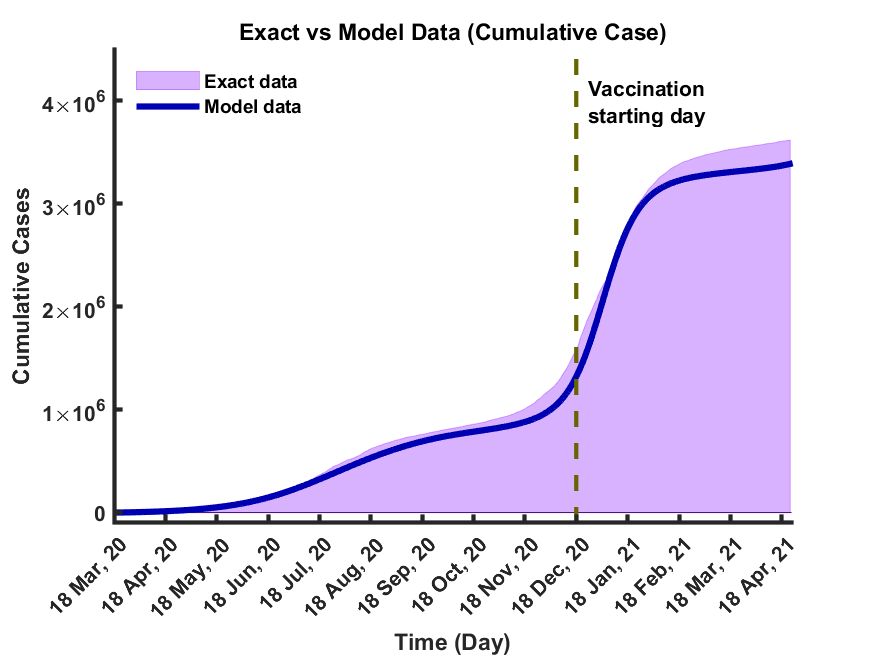}
	\caption{Model plausibility check for SARS-CoV-2 pandemic: Daily reported cases (left), and cumulative cases (eight) with `Pfizer' vaccine.}
	\label{case_fit}
\end{figure}

\begin{figure} [H]
	\centering
	\includegraphics[width=0.48 \linewidth]{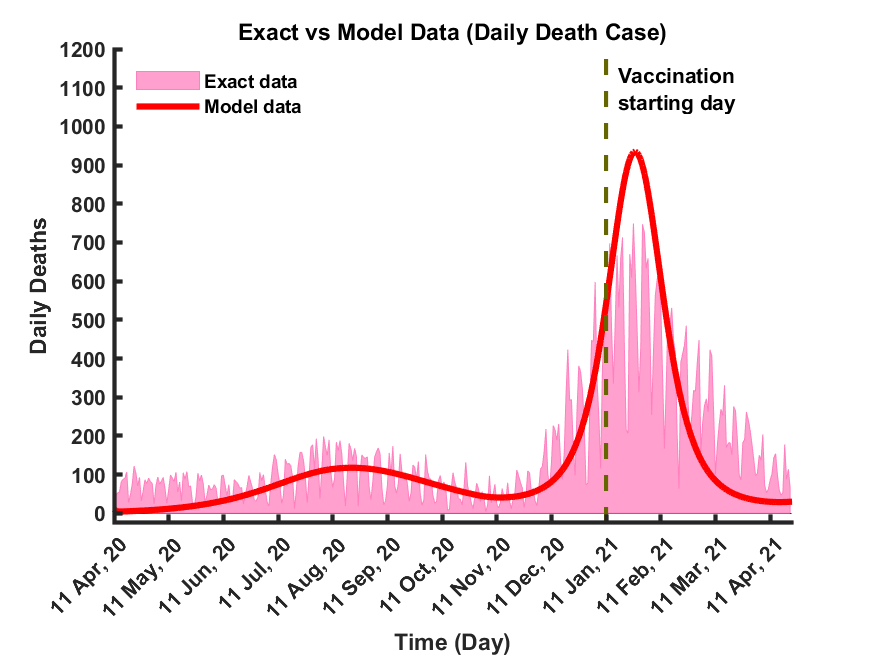}
	\includegraphics[width=0.48 \linewidth]{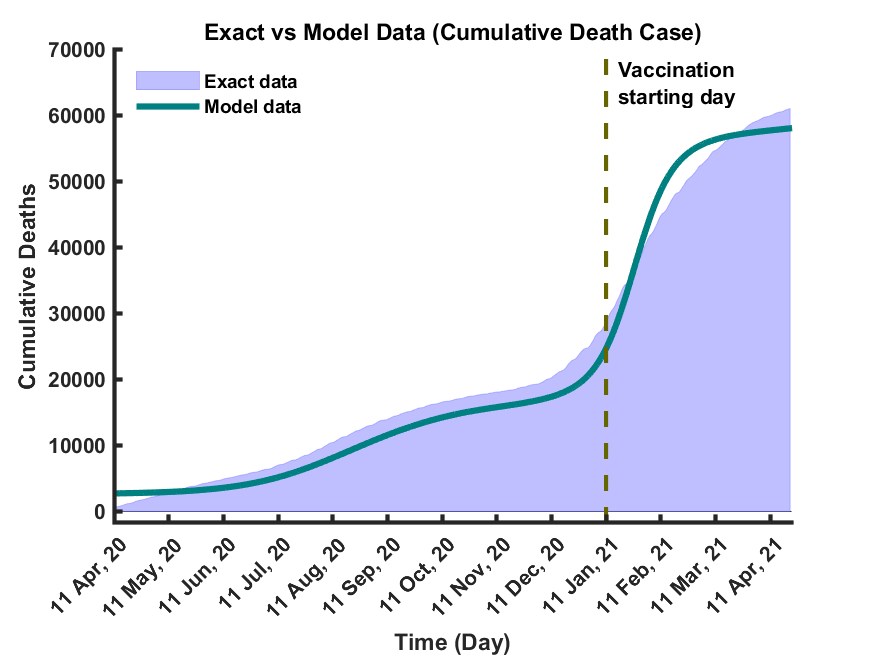}
	\caption{Model plausibility check for SARS-CoV-2 pandemic: Daily reported death cases (left), and cumulative death cases (right) with `Pfizer' vaccine.}
	\label{death_fit}
\end{figure}

Figure \ref{case_fit} illustrates the plausibility fit of the considered model for the available SARS-CoV-2 data for California with the usage of the `Pfizer' vaccine from December 15, 2021, to April 17, 2021. From the actual data of daily cases, it is clear that, till April 2021, there have been two waves of infection in California. Moreover, the figures in Figure \ref{death_fit} depict the model fitting with the actual fatality data in California. To fit the wave data, we have induced the concept of viral mutation as a piece-wise parameter value with an average intervention of 20 weeks. Furthermore, this model data forecasts several waves upwards.

\begin{figure} [H]
	\centering
	\includegraphics[width=0.8 \linewidth]{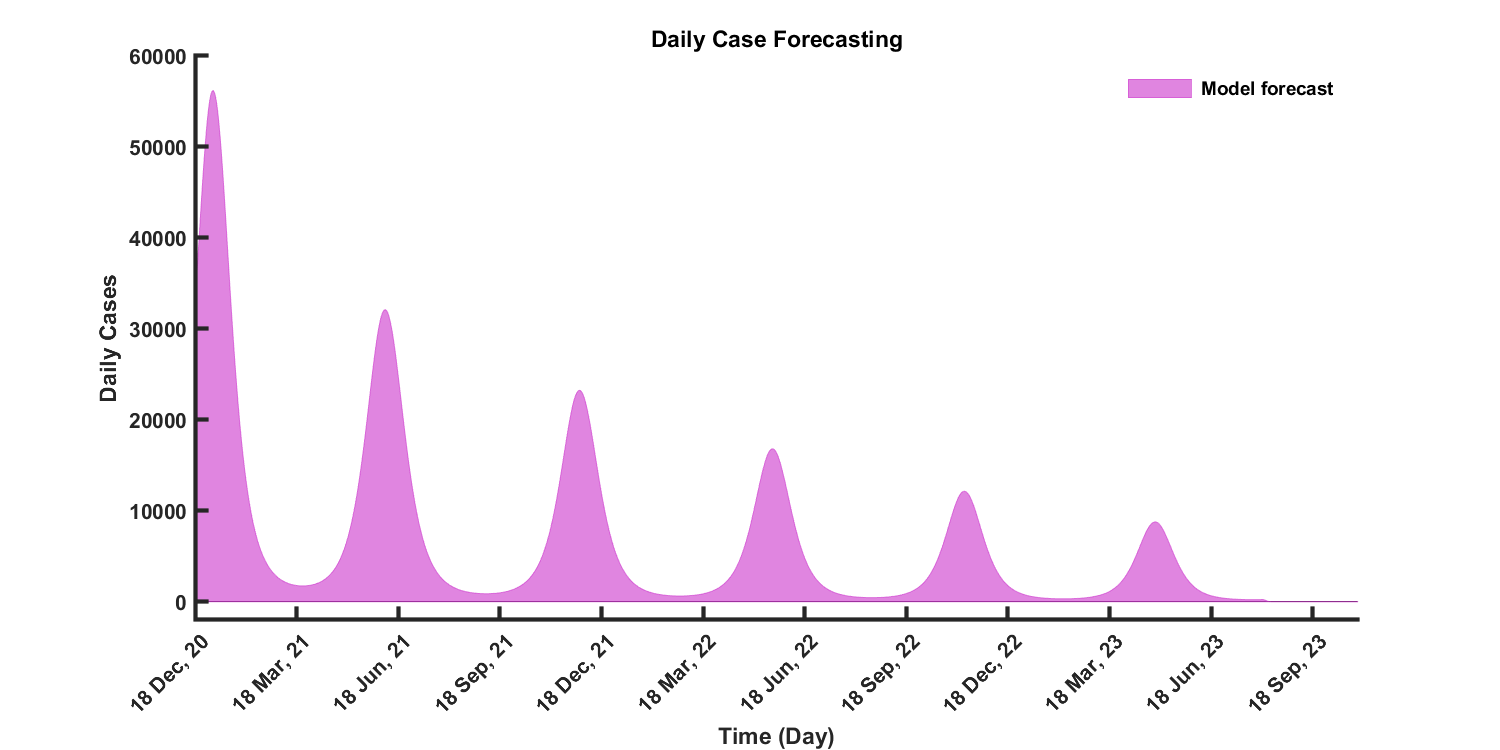}
	\caption{Model forecasting: Daily confirmed cases %(above), and Cumulative cases (below)
		with `Pfizer' vaccine.}
	\label{case_fore}
\end{figure}

\begin{figure} [H]
	\centering
	\includegraphics[width=0.8 \linewidth]{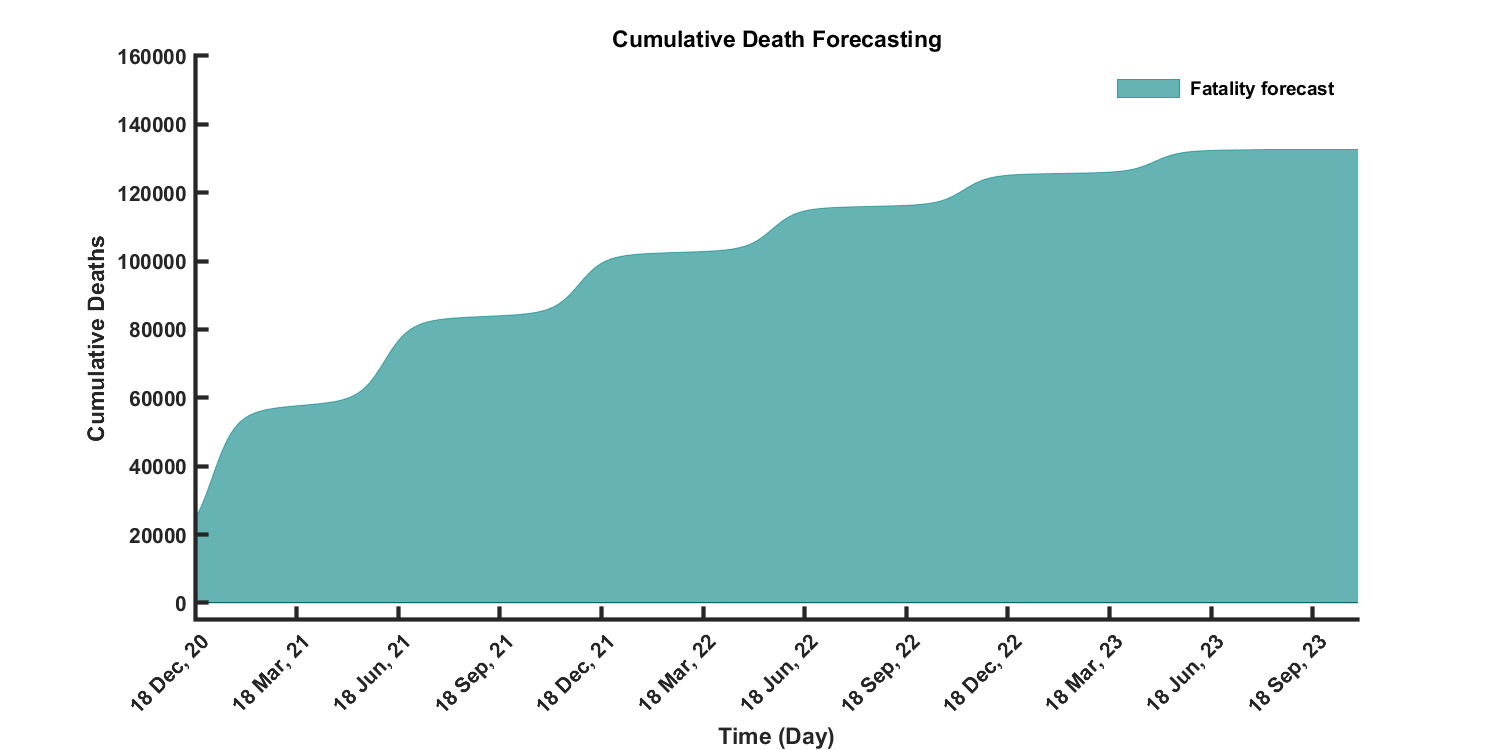}
	\caption{Model forecasting: Cumulative death cases %(above), and Cumulative cases (below)
		with `Pfizer' vaccine.}
	\label{death_fore}
\end{figure}

Figure \ref{case_fore} shows all the possible upcoming waves of SARS-CoV-2 in California, U.S., even if the people are continuously being vaccinated with the Pfizer vaccine, whose effectiveness rate is $ 97\% $ against the symptomatic COVID-19 cases, hospitalizations, severe and critical hospitalizations, and deaths (94\% against asymptomatic SARS-CoV-2 infections) even in case of the full vaccine dose completion \cite{Pfizer}. The model forecasts 6 (six) waves in total from the day of the vaccination program started in this state, December 15, 2021 (seven waves during the total pandemic period). It is also observable that the wave peaks are reducing as time passes, and the epidemic seems to be under control by the mid of 2023. In the meanwhile, 7,814,589 cases will be confirmed according to this study, with 132,636 deaths (Figure \ref{death_fore}) which is 1.697\% of a total case with 98.30\% recovered patients. Till the mid of 2023, about 20\% of the total population of California will be infected with SARS-CoV-2.

%\begin{figure} [H]
%	\centering
%	\includegraphics[width=0.55 \linewidth]{Codes/cal_s_multi.png}
%	\includegraphics[width=0.4 \linewidth]{Codes/cal_s_multi-1.png}
%\end{figure}
%\begin{figure} [H]
%	\centering
%	\includegraphics[width=0.55 \linewidth]{Codes/cal_v_multi.png}
%	\includegraphics[width=0.4 \linewidth]{Codes/cal_v_multi-1.png}
%\end{figure}
%\begin{figure} [H]
%	\centering
%	\includegraphics[width=0.55 \linewidth]{Codes/cal_e_multi.png}
%	\includegraphics[width=0.4 \linewidth]{Codes/cal_e_multi-1.png}
%\end{figure}
%\begin{figure} [H]
%	\centering
%	\includegraphics[width=0.55 \linewidth]{Codes/cal_i_multi.png}
%	\includegraphics[width=0.4 \linewidth]{Codes/cal_i_multi-1.png}
%\end{figure}
%\begin{figure} [H]
%	\centering
%	\includegraphics[width=0.55 \linewidth]{Codes/cal_r_multi.png}
%	\includegraphics[width=0.4 \linewidth]{Codes/cal_r_multi-1.png}
%	\caption{Model forecasting against all available vaccines.}
%	\label{com_fore}
%\end{figure}

\begin{figure} [H]
	\centering
	\subfloat[\label{f60}]{\includegraphics[width = 0.55 \linewidth]{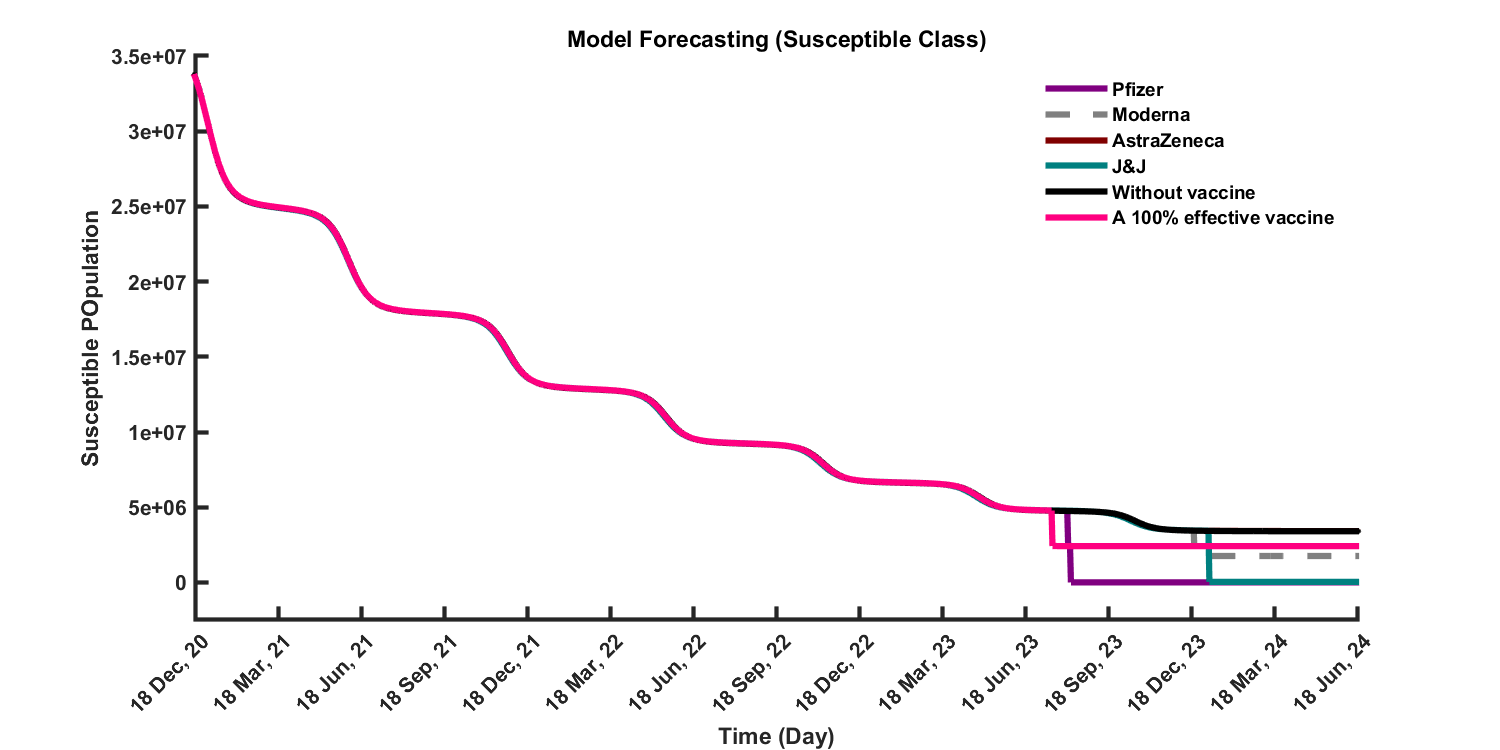}}
	\subfloat[\label{f61}]{\includegraphics[width = 0.5 \linewidth]{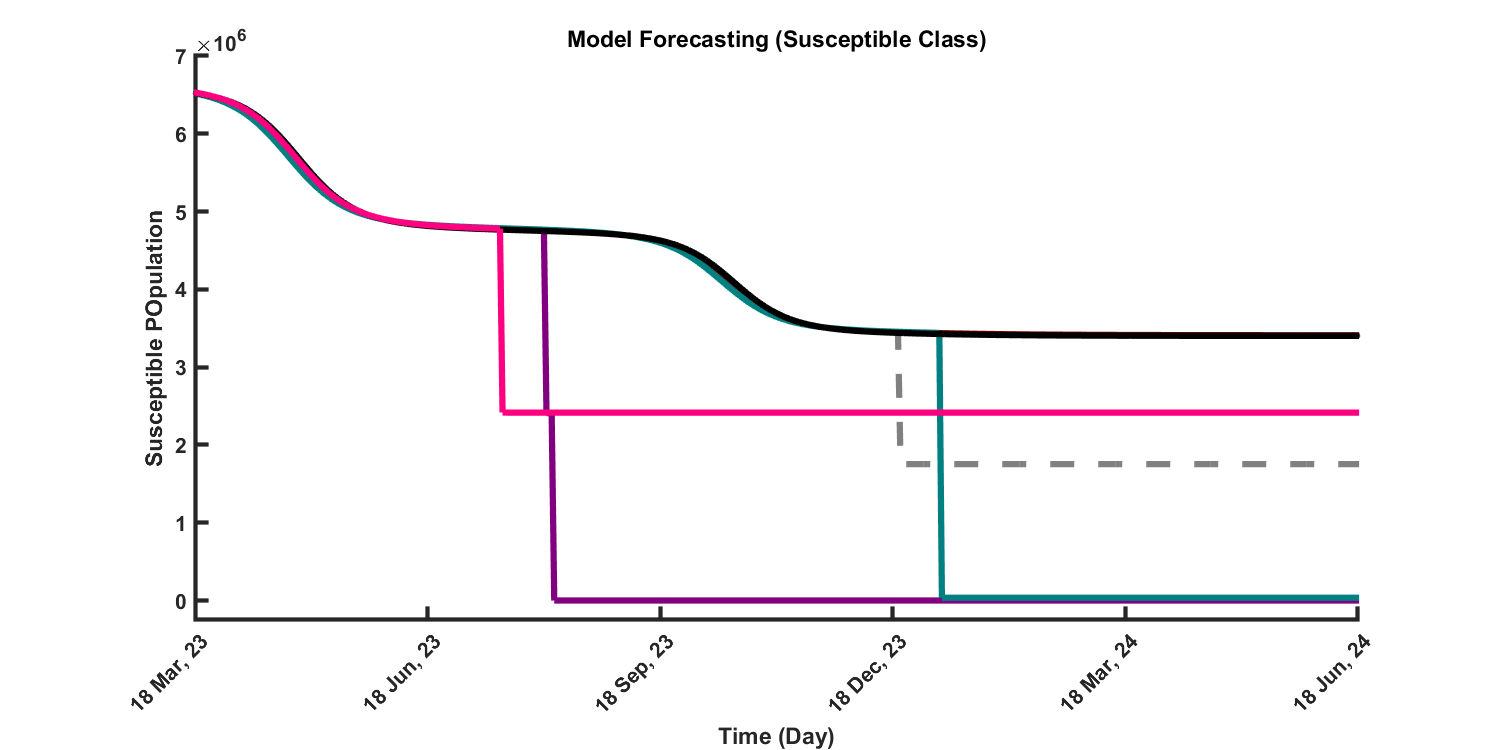}} \\
\end{figure}

\begin{figure} [H] \ContinuedFloat
	\centering
	\subfloat[\label{f62}]{\includegraphics[width = 0.55 \linewidth]{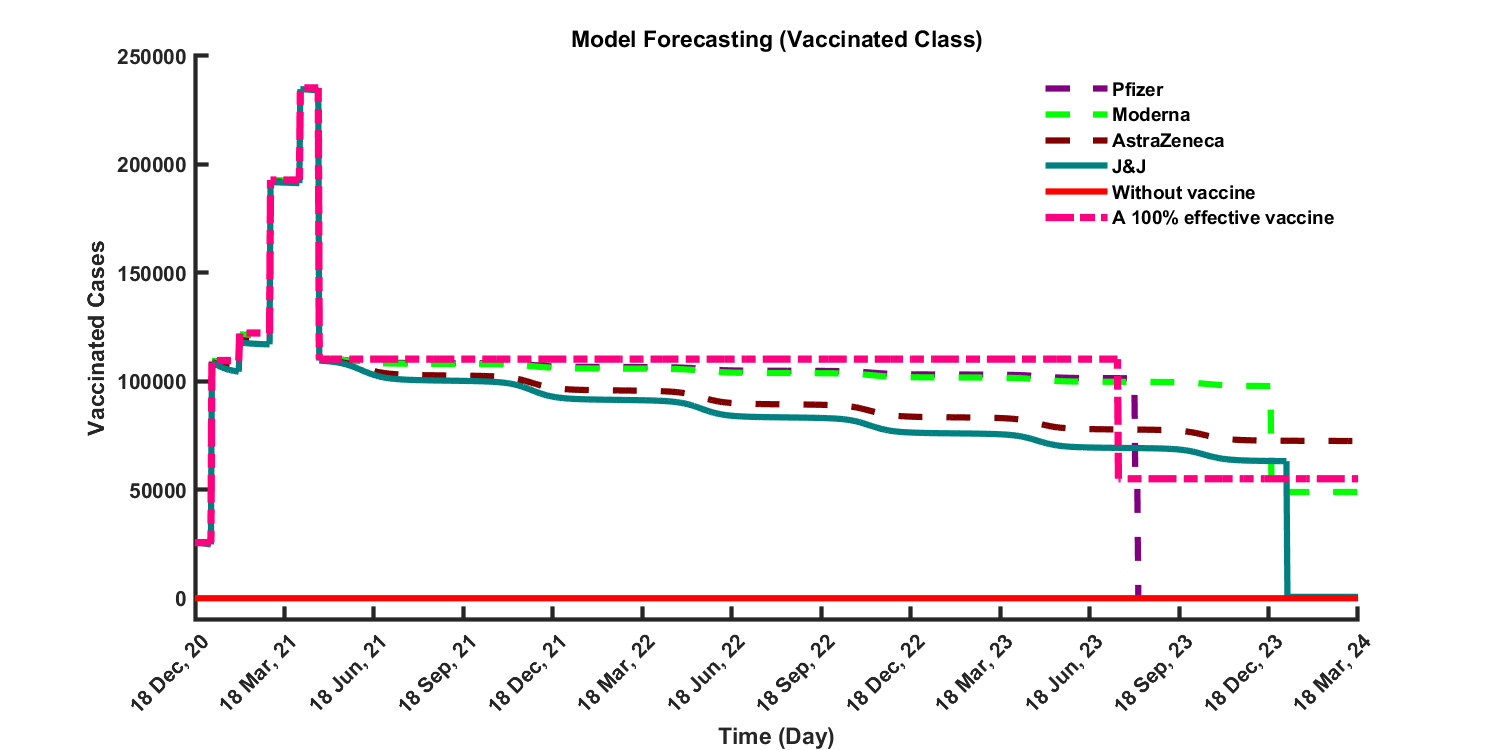}}
	\subfloat[\label{f63}]{\includegraphics[width = 0.5 \linewidth]{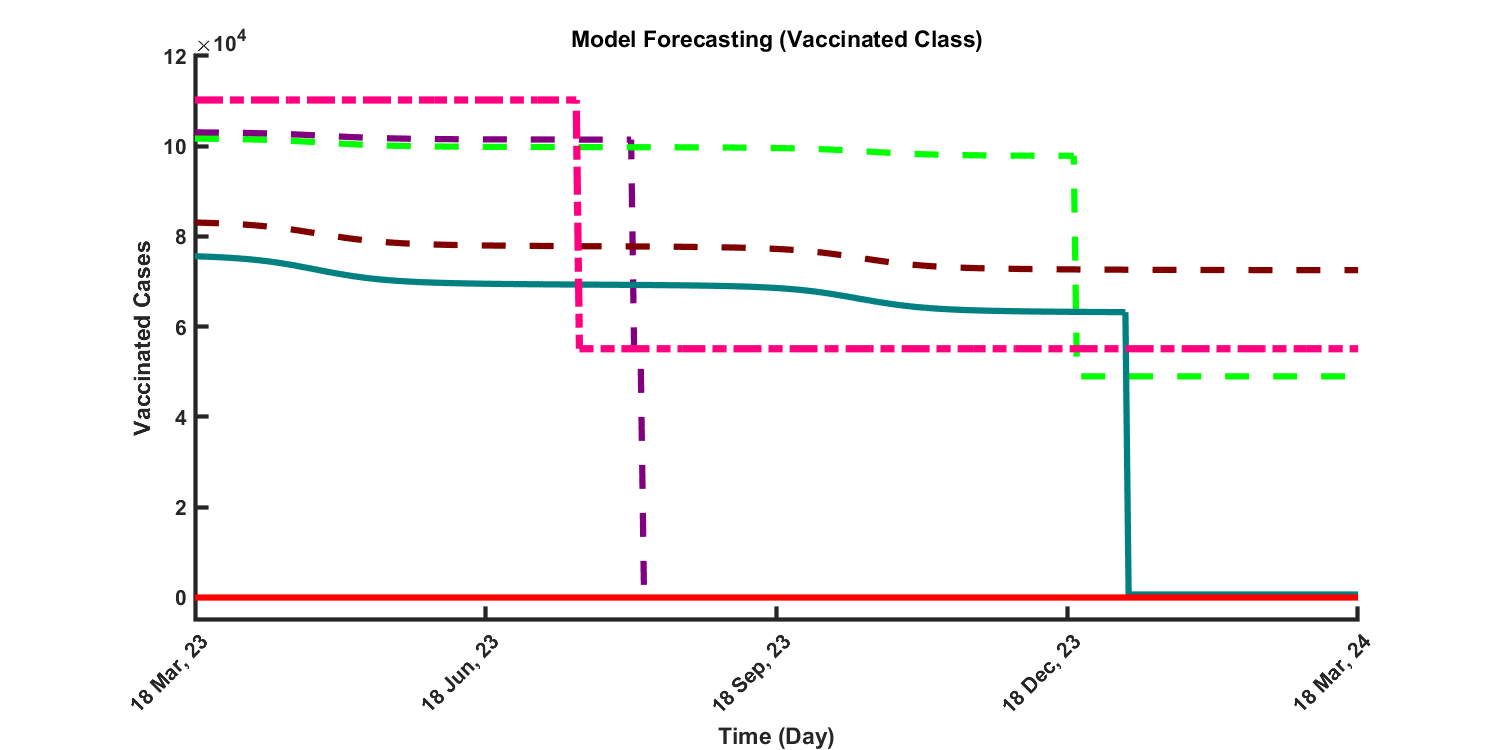}} \\
\end{figure}

\begin{figure} [H] \ContinuedFloat
	\centering
	\subfloat[\label{f64}]{\includegraphics[width = 0.55 \linewidth]{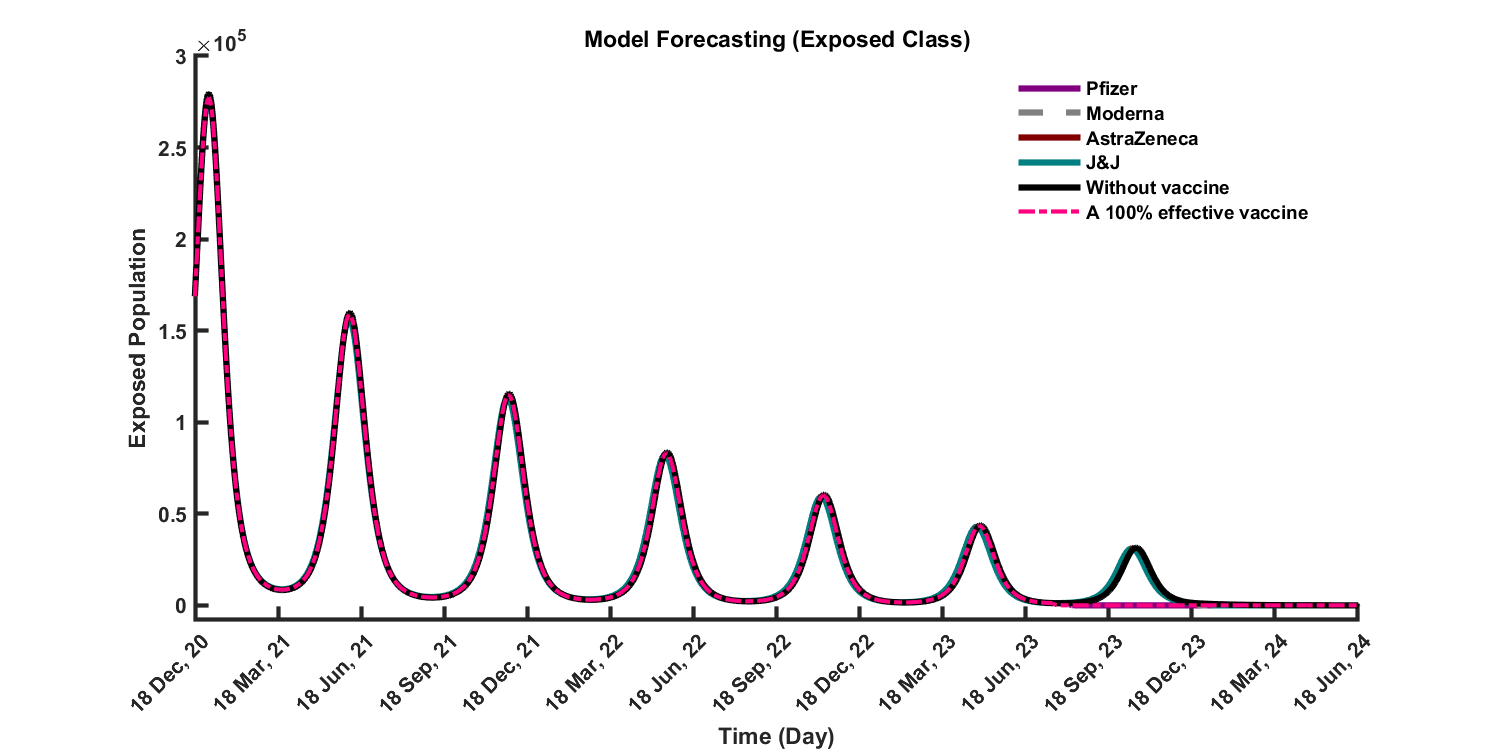}}
	\subfloat[\label{f65}]{\includegraphics[width = 0.5 \linewidth]{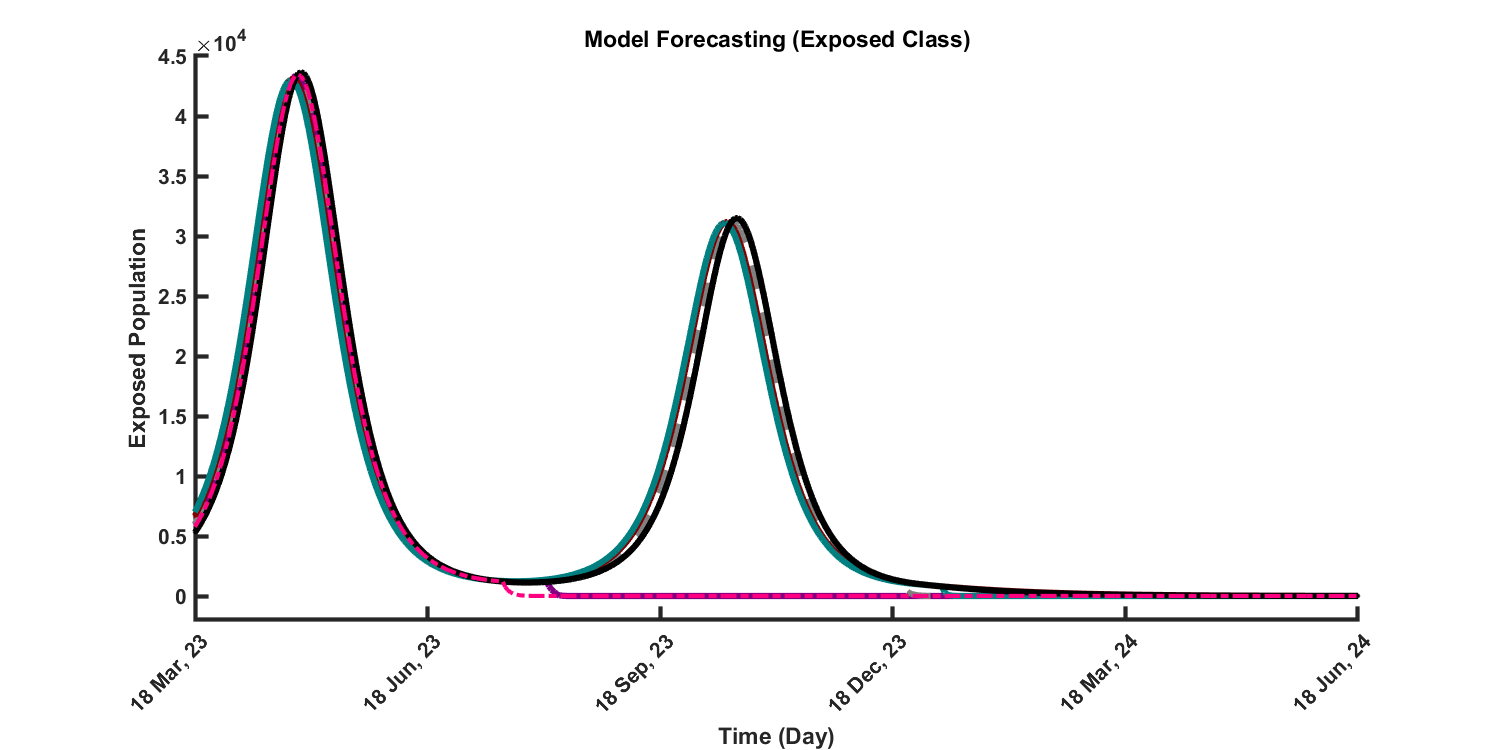}} \\
\end{figure}

\begin{figure} [H] \ContinuedFloat
	\centering
	\subfloat[\label{f66}]{\includegraphics[width = 0.55 \linewidth]{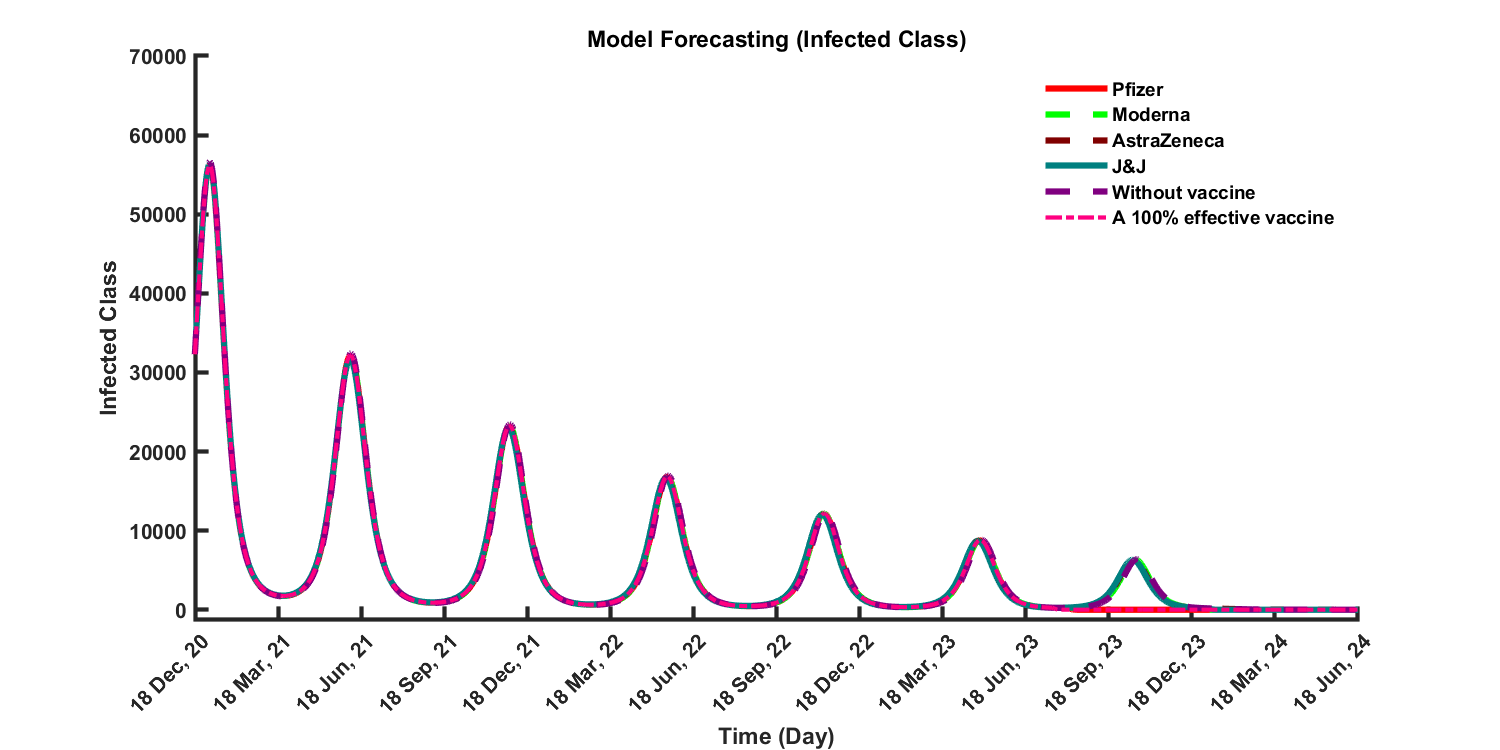}}
	\subfloat[\label{f67}]{\includegraphics[width = 0.5 \linewidth]{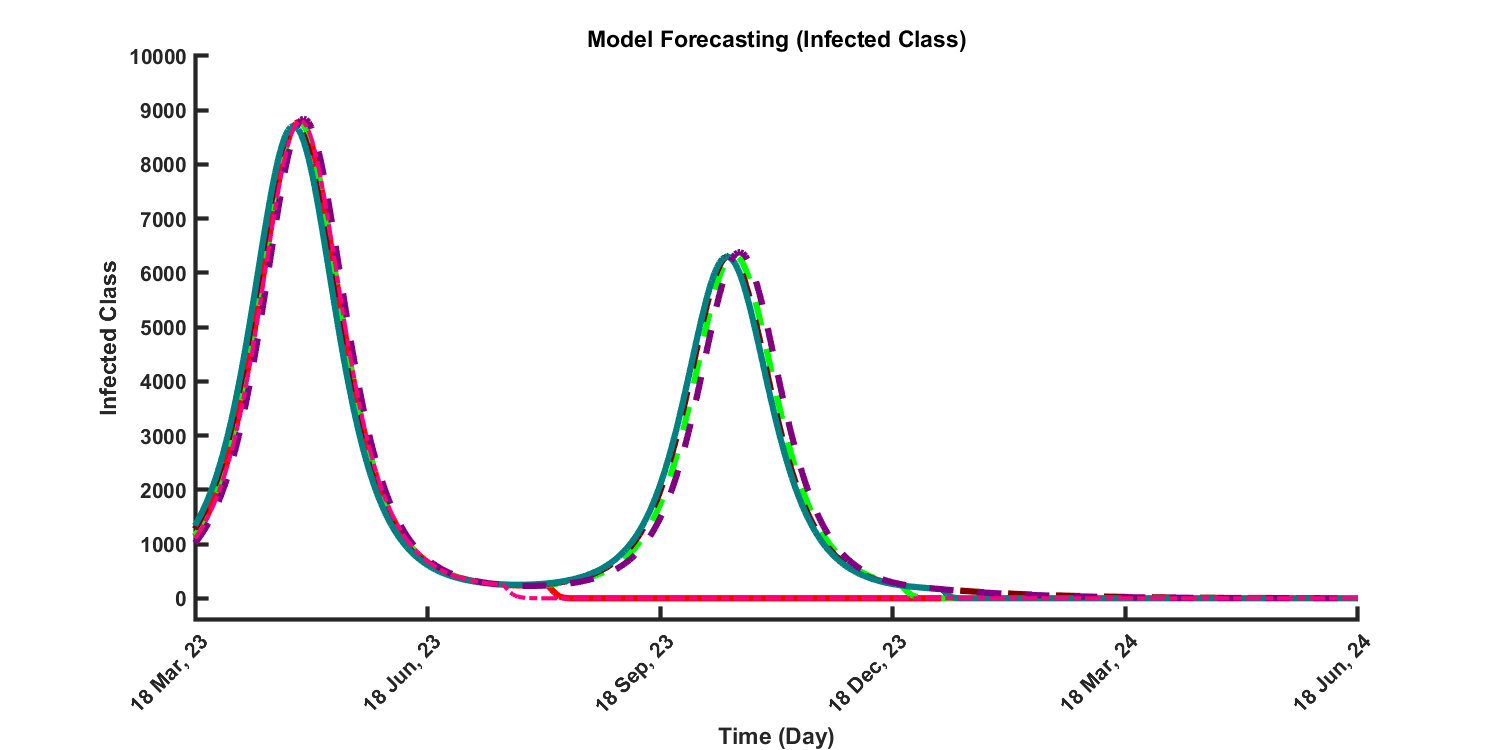}} \\
\end{figure}

\begin{figure} [H] \ContinuedFloat
	\centering
	\subfloat[\label{f68}]{\includegraphics[width = 0.55 \linewidth]{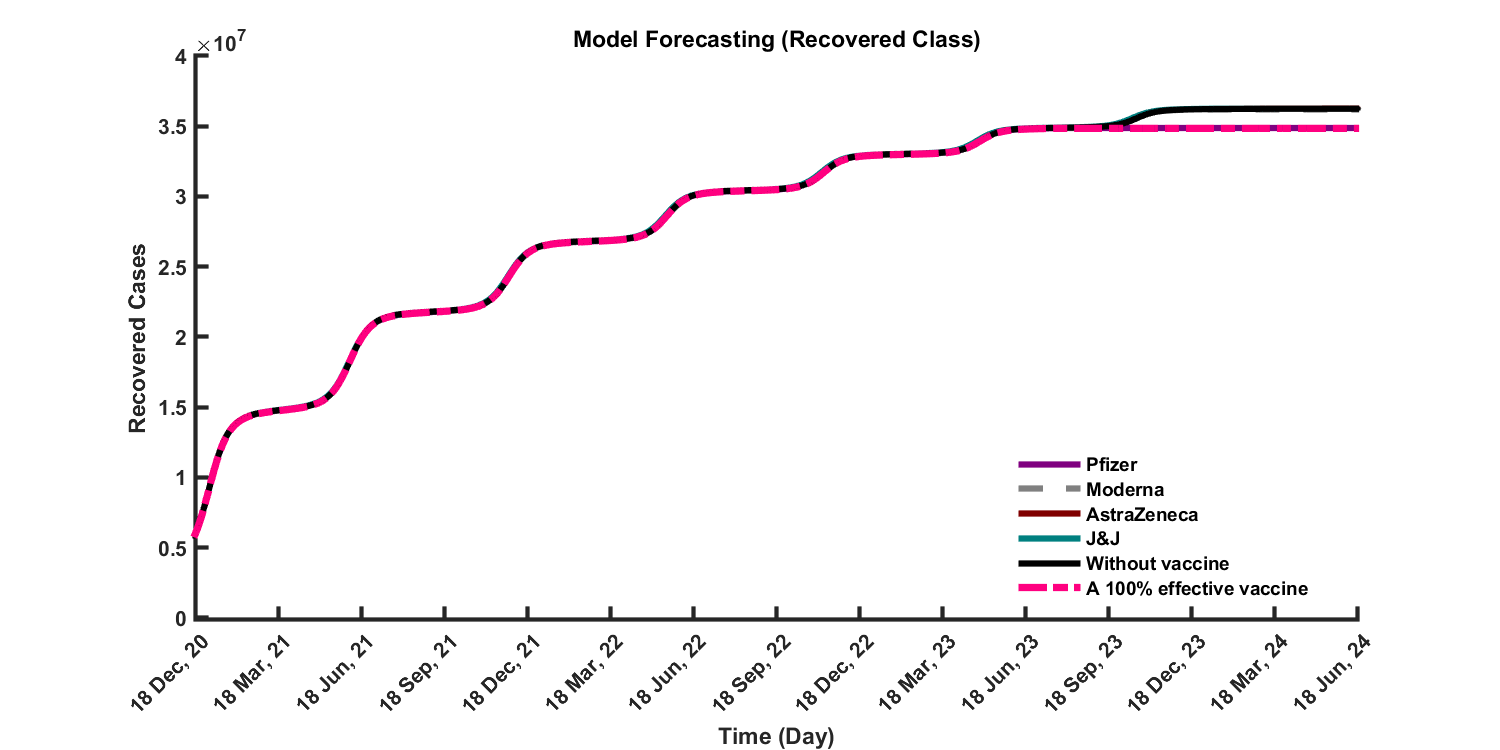}}
	\subfloat[\label{f69}]{\includegraphics[width = 0.5 \linewidth]{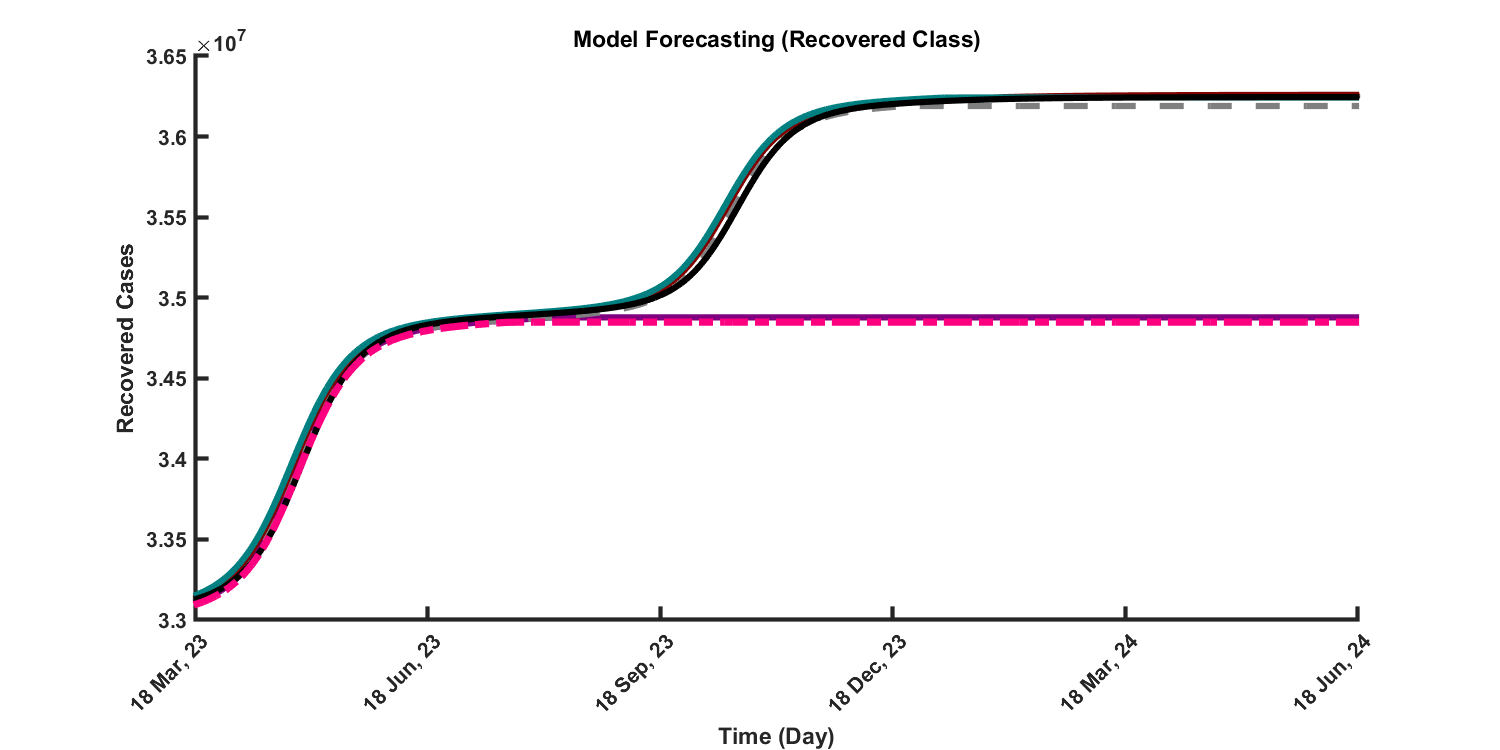}} \\
	\caption{Model forecasting against all available vaccines.}
	\label{com_fore}
\end{figure}

Until eradicating the infection as a pandemic, mid-2023, every considered compartment of the model gets some rigorous changes corresponding to different vaccine scenarios: Pfizer, Moderna, AstraZeneca, Johnson \& Johnson, without any vaccine, and even in the implication of a 100\% effective vaccine. The reduction of the susceptible compartment is noticeable (Figure \ref{f60}, and \ref{f61}). Since the vaccination rate in California has not been very consistent, we have calculated the average vaccination count per month and proceeded for the vaccination compartment, for which this compartment shows its daily changes step by step and reduces to the null point when the pandemic is over (Figure \ref{f62}, and \ref{f63}).

The effects of vaccines for the $ E $ and $ I $ classes are also shown in the Figure (\ref{f64}, \ref{f65}, \ref{f66}, and \ref{f67}). Every wave is much smaller than the previous wave peak for all vaccine scenarios, and the classes vanish gradually, which means the end of the SARS-CoV-2 pandemic. Only the Pfizer vaccine (97\% effectiveness) controls the pandemic one wave earlier than the other vaccines; when a hypothetical presence of a 100\% effective vaccine controls before Pfizer. Surprisingly, other vaccines whose effectiveness rates are below 95\% cannot make much difference from the `no vaccine' scenario. The same cases are shown for the recovered class in the Figure (\ref{f68}, and \ref{f69}).

\begin{figure} [H]
%	\fbox{\begin{minipage}{6.25in}
			\centering
			\subfloat[\label{f35}]{\includegraphics[width = 0.42 \linewidth]{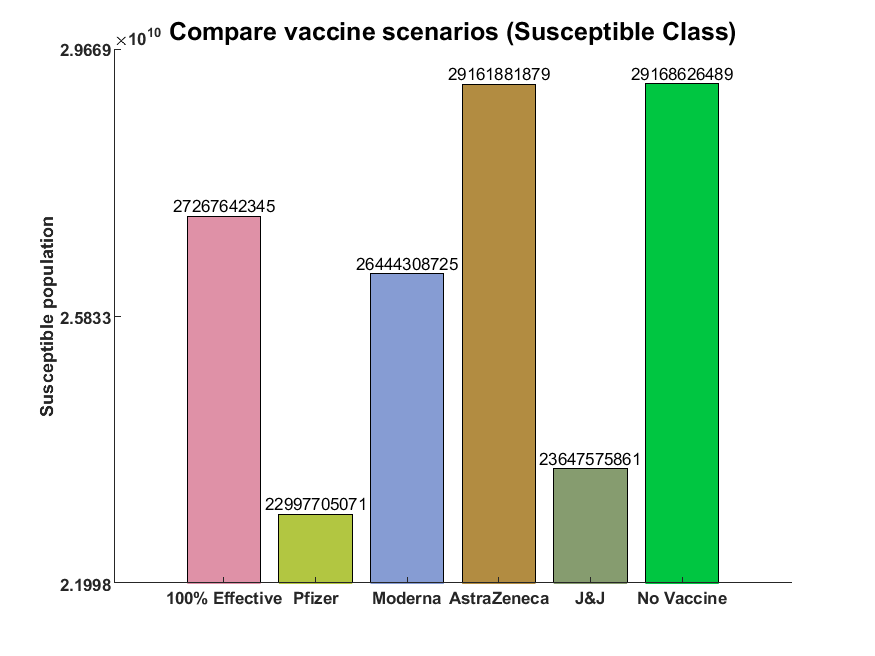}}
			\subfloat[\label{f36}]{\includegraphics[width = 0.42 \linewidth]{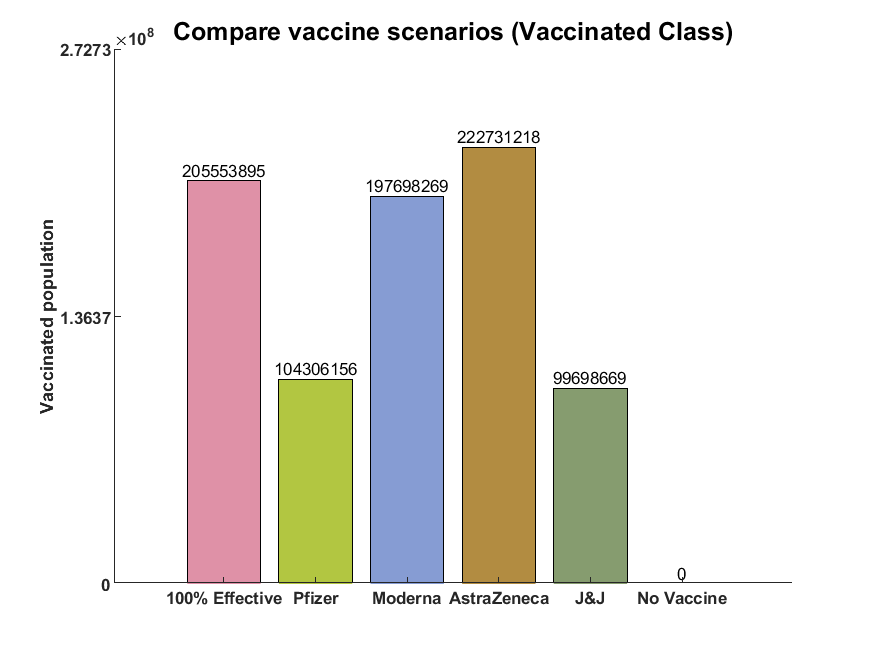}} \\
			\subfloat[\label{f37}]{\includegraphics[width = 0.42 \linewidth]{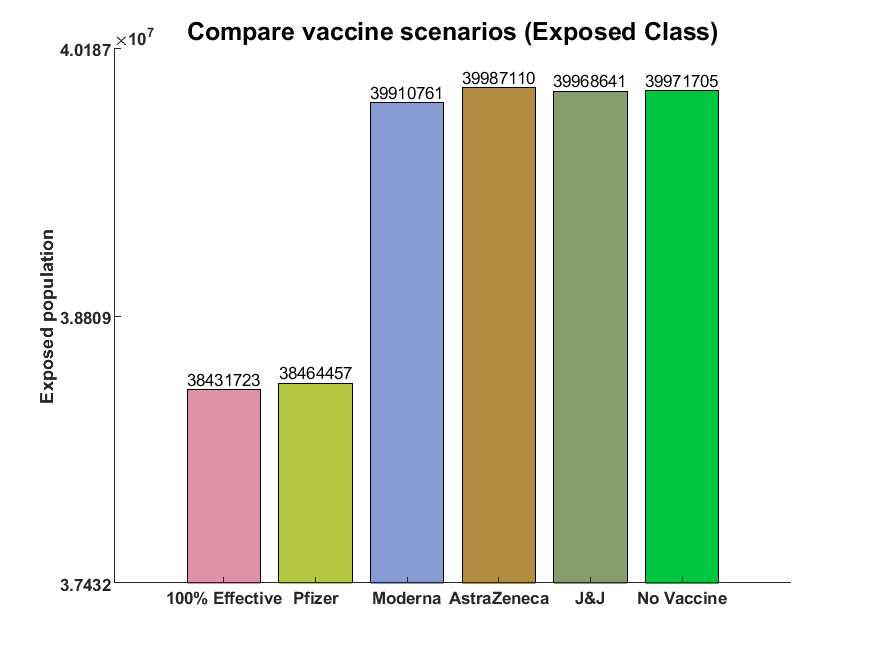}}
			\subfloat[\label{f38}]{\includegraphics[width = 0.42 \linewidth]{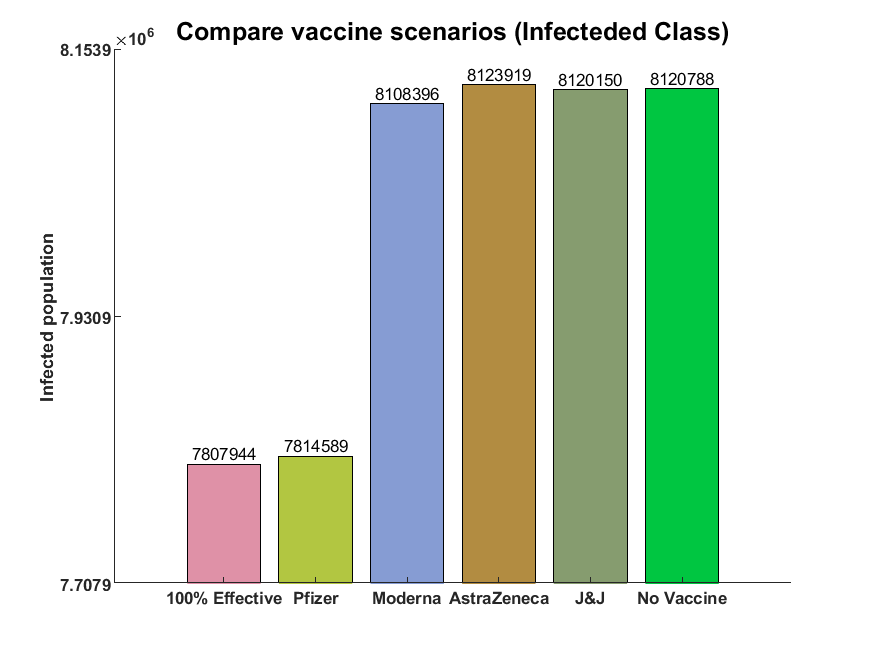}} \\
			\subfloat[\label{f39}]{\includegraphics[width = 0.42 \linewidth]{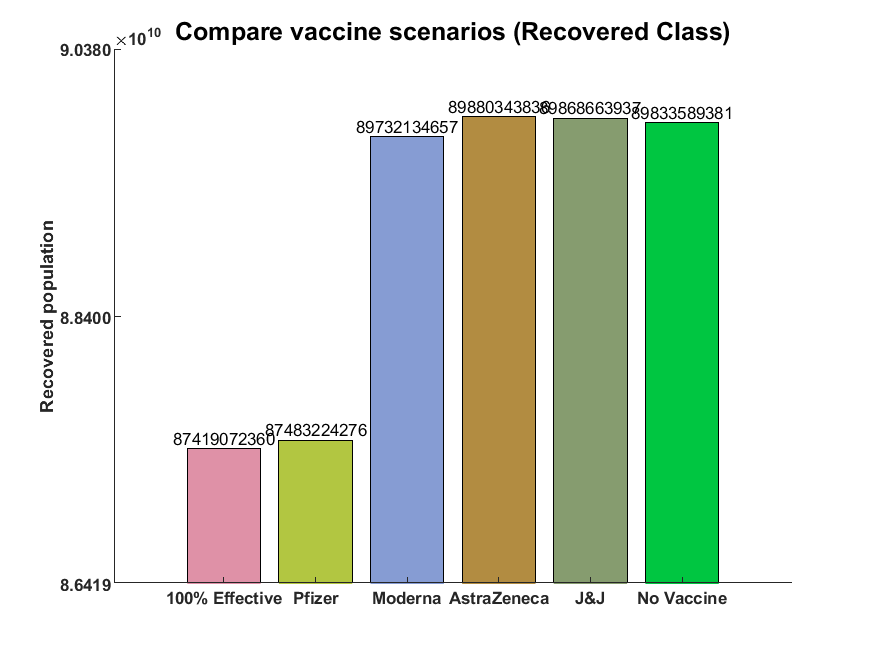}}
%	\end{minipage}}
	\caption{Effects of different vaccine situations in California, U.S.}
	\label{fig7}
\end{figure}

Figure \ref{fig7} depicts compare among all vaccine scenarios at the foretasted end of the pandemic situations for all compartments (Susceptible, Vaccinated, Exposed, Infected, and Recovered). The numerical study also takes into account that when the number of active confirmed cases decreases down to less than 100 in the whole state (California) and remains less than 100 active cases for more than 20 weeks, then the pandemic will be considered as under control and going to be eradicated forever. As a result, after that moment- the vaccination program will be no longer necessary and so will be shut down step-by-step amidst the next 5-6 weeks.

\begin{figure} [H]
	\centering
	\includegraphics[width=0.48 \linewidth]{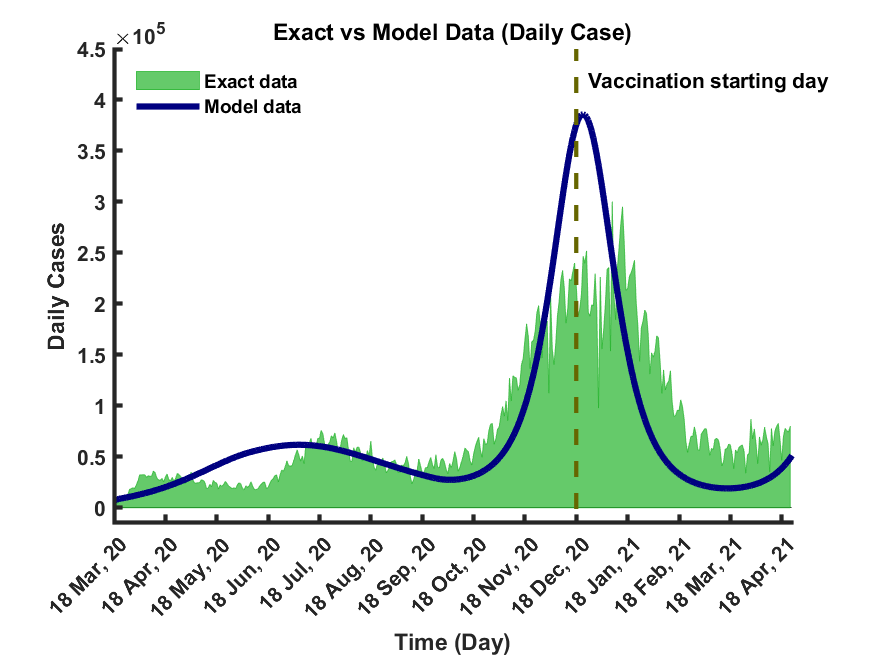}
	\includegraphics[width=0.48 \linewidth]{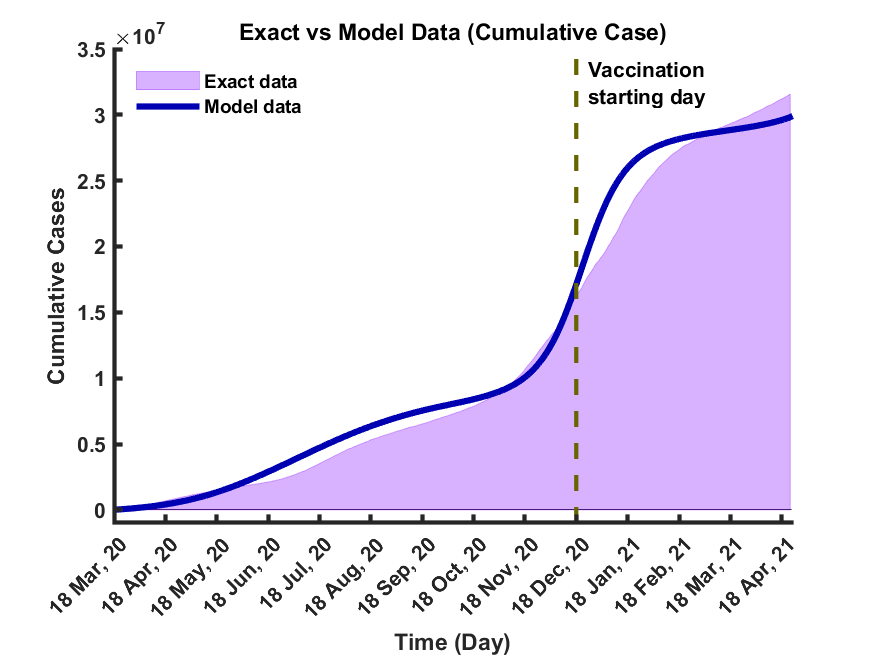}
	\caption{Model data fitting to U.S. SARS-CoV-2 data.}
	\label{case_fit_us}
\end{figure}

\begin{figure} [H]
	\centering
	\includegraphics[width=0.8 \linewidth]{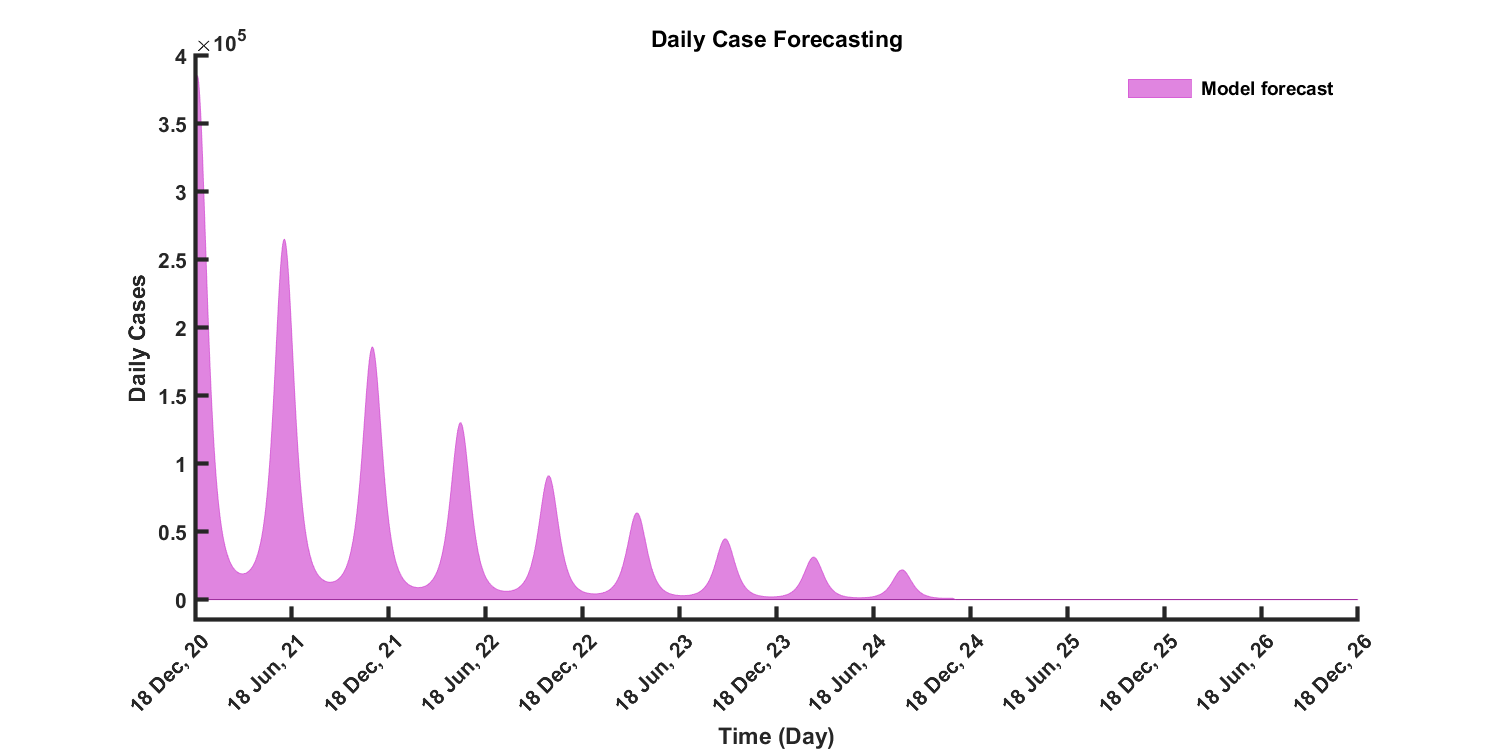}
	\caption{Model forecasting to U.S. SARS-CoV-2 situation.}
	\label{case_fore_us}
\end{figure}

\begin{figure} [H]
	\centering
	\includegraphics[width=0.8 \linewidth]{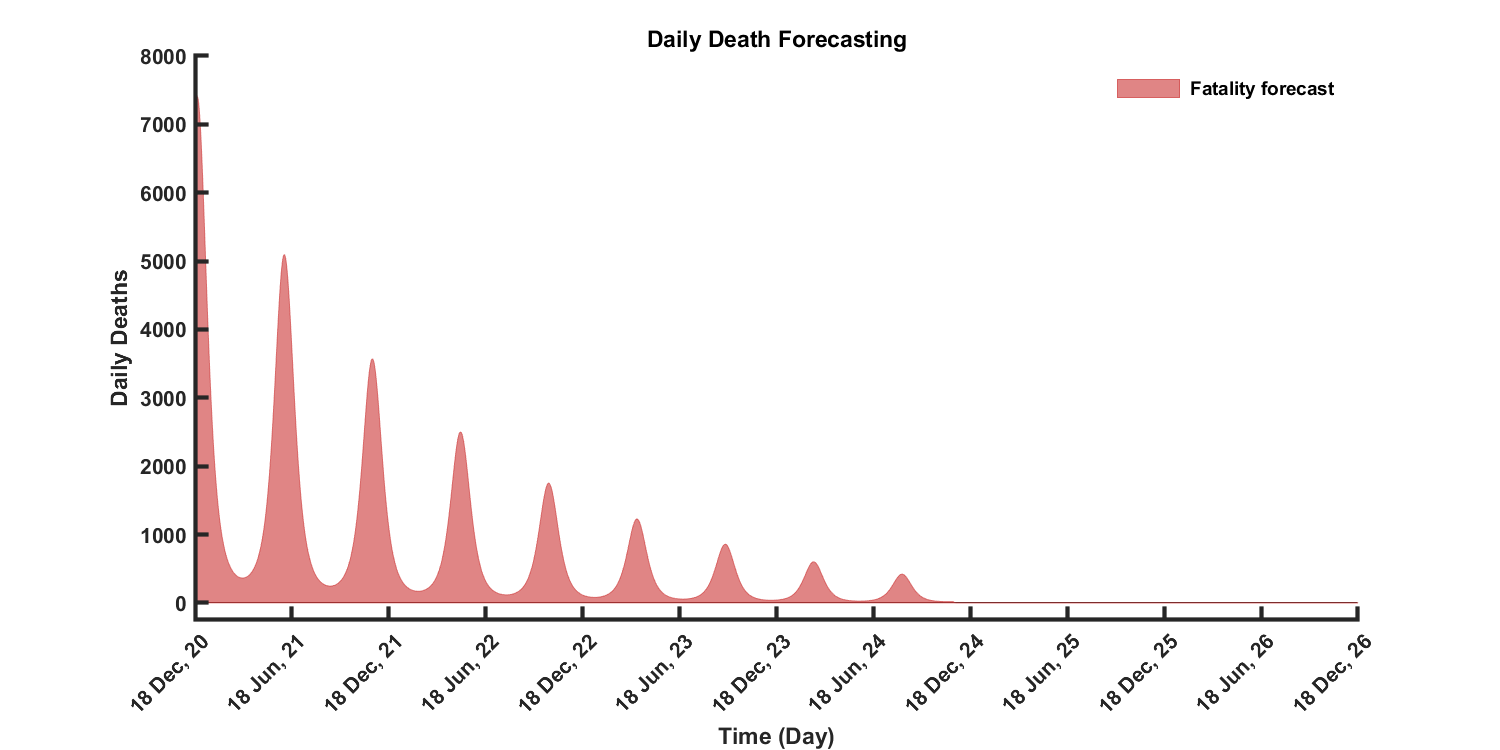}
	\caption{Model forecasting to U.S. SARS-CoV-2 fatalities.}
	\label{death_fore_us}
\end{figure}

The model has also been simulated for the SARS-CoV-2 data for the United States of America in the Appendix A.% \ref{appendix-a}. 
The model is well fitted for the real data of the U.S. (Figure \ref{case_fit_us}). There will be seven to eight big waves of infection from December 2020 to December 2023 and five to six minor waves afterward, and the pandemic is gone entirely by the end of 2026 (Figure \ref{case_fore_us}, and \ref{death_fore_us}). According to this study, 73,766,040 infection cases and 1,442,772 death cases may get reported at the end of 2026 in the U.S.

The compartmental dynamics for different vaccine scenarios has been shown in the Figure \ref{com_fore_us}. The sub-figures \ref{fu0},\ref{fu2},\ref{fu4},\ref{fu6}, and \ref{fu8} depicts the gradual change of the corresponding compartment up to the stable situation from the December 2020, but since it's being difficult to visualize the vaccine related variation the sub-figures \ref{fu1},\ref{fu3},\ref{fu5},\ref{fu7}, and \ref{fu9} displays the same results from June 2024 to June 2028 where the crucial changes are happening due to the different vaccine efficacy rates for the Susceptible-Vaccinated-Exposed-Infected-Recovered compartments, respectively.

\section*{Findings \& decisions}
Like many other epidemics, the vaccine may pull down the death tolls by SARS-CoV-2. There is no way to erase this infectious disease overnight; rather, we have to be used to this virus and live our everyday life with all prescribed precautions, not only as long as we are in this pandemic, but for the rest of our lives to maintain a hygienic and healthy life, and to make sure another pandemic is never coming. Besides all these lifestyles, everyone must be involved in the vaccination program as soon as possible.

Our study finds that a 100\% clinically feasible vaccine for all variants of this virus is necessary to reduce the disease-induced loss. The vaccines which are less than 95\% adequate should not be recommended, as they will make no good than the cost of both time and resources.

Since the vaccine efficacy rates for all these considered vaccines are not strict per virus variants, physical nature/condition of people from different localities and ethnicity, vaccine allocation is not perfect, and many more \cite{Pfizer,astra,moderna,allocation}, the efficacy rates we have considered throughout this study resembles any other kind of vaccines which are of the same efficacy rates of the mentioned vaccines with a maximum deviation of 2-3\%.

\section*{Concluding remarks}
There are no other effective ways to fight against any pandemic diseases except vaccination. Although specific groups of people get prioritized initially, mass vaccination is needed to control the spread of the disease. The world is still suffering due to the COVID-19 pandemic that started more than one and a half years ago. Several vaccines with different efficacy have been used globally; infection is still on the rise in many countries. It is hard to predict when a pandemic will be eradicated. In this mathematical modeling study, we used California, and U.S. data, and four vaccines to forecast the possible upcoming COVID-19 situations in the U.S.. Since the considered vaccines have different efficacy rates, we have also demonstrated the impact of efficacy rates on containing the pandemic.

\section*{Acknowledgment}\label{acknow}
The research by M. Kamrujjaman was partially supported by a TWAS grant 2019\_19-169 RG/MATHS/AS\_I.
%The authors M. Kamrujjaman and M. S. Islam are grateful to the University of Dhaka, the 100 years centennial research program. 
%The authors are grateful to Professor Dr. Taufiquar Khan for his constructive suggestions on the primary version of the manuscript.

%\section*{Funding statement}
%The research by M. Kamrujjaman was partially supported by a TWAS grant 2019\_19-169 RG/MATHS/AS\_I.
%The authors  M. Kamrujjaman and M. Shahidul Islam research were partially supported by University of Dhaka Centennial Research Grant.

\section*{Conflict of interest}
The authors declare no conflict of interest exists.

\section*{Author contributions}
Conceptualization, M.K. and M.S.M.; methodology, M.S.M., M.M.I.Y.A and M.A.H; software, M.S.M., M.M.I.Y.A and M.K.; validation, M.A.H, M.M.R., and M.M.; formal analysis, M.S.M., M.K. and M.A.H; investigation, M.S.I; resources, M.M.M. and M.M.R; data curation, M.S.M., M.M.I.Y.A and M.A.H; original draft preparation, M.S.M., M.K., M.M.I.Y.A and M.A.H; review and editing, M.M.M., M.M.R., M.M. and M.S.I; supervision, M.K. All authors have read and agreed to the published version of the manuscript.

\section*{Disclaimer/data availability statement}
 All data are provisional and subject to change. Probable cases are
not included in the total case numbers. No consent is required to publish this paper

%
%\newpage
%\section*{\rm Glossary of Notation}
%\begin{tabbing}
%	\hspace{2.0cm}\=\hspace{4in}\=\kill
%	
%	$ \Omega $ \> Bounded spatial habitat \\
%	$ \partial\Omega $ \> Smooth boundary of bounded spatial habitat $ \Omega $ \\
%	$ \mathbb{R} $ \> Set of real numbers \\
%	$ \mathbb{R}^{n} $ \> Set of ordered $ n $-tuples of real numbers \\
%	$ \mathcal{R}_{0} $ \> Basic reproduction number \\
%	$ E_{0} $ \> Disease-free equilibrium \\
%	$ E^{*} $ \> Disease equilibrium \\
%\end{tabbing}

%\newpage

%\clearpage
\section*{Appendix A} \label{appendix-a}

\section*{Case: The United States of America}

%\begin{figure} [H]
%	\centering
%	\includegraphics[width=0.55 \linewidth]{Codes/usa_s_multi.png}
%	\includegraphics[width=0.4 \linewidth]{Codes/usa_s_multi_1.png}
%	\caption{Model forecasting against all available vaccines: Susceptible compartment.}
%	\label{com_fore_us_s}
%\end{figure}
%
%\begin{figure} [H]
%	\centering
%	\includegraphics[width=0.55 \linewidth]{Codes/usa_v_multi.png}
%	\includegraphics[width=0.4 \linewidth]{Codes/usa_v_multi_1.png}
%	\caption{Model forecasting against all available vaccines.}
%	\caption{Model forecasting against all available vaccines: Vaccinated compartment.}
%	\label{com_fore_us_v}
%\end{figure}
%
%\begin{figure} [H]
%	\centering
%	\includegraphics[width=0.55 \linewidth]{Codes/usa_e_multi.png}
%	\includegraphics[width=0.4 \linewidth]{Codes/usa_e_multi_1.png}
%	\caption{Model forecasting against all available vaccines: Exposed compartment.}
%	\label{com_fore_us_e}
%\end{figure}
%
%\begin{figure} [H]
%	\centering
%	\includegraphics[width=0.55 \linewidth]{Codes/usa_i_multi.png}
%	\includegraphics[width=0.4 \linewidth]{Codes/usa_i_multi_1.png}
%	\caption{Model forecasting against all available vaccines: Infected compartment.}
%	\label{com_fore_us_i}
%\end{figure}
%
%\begin{figure} [H]
%	\centering
%	\includegraphics[width=0.55 \linewidth]{Codes/usa_r_multi.png}
%	\includegraphics[width=0.4 \linewidth]{Codes/usa_r_multi_1.png}
%	\caption{Model forecasting against all available vaccines: Recovered compartment.}
%	\label{com_fore_us_r}
%\end{figure}

\begin{figure} [H]
	\centering
	\subfloat[\label{fu0}]{\includegraphics[width = 0.55 \linewidth]{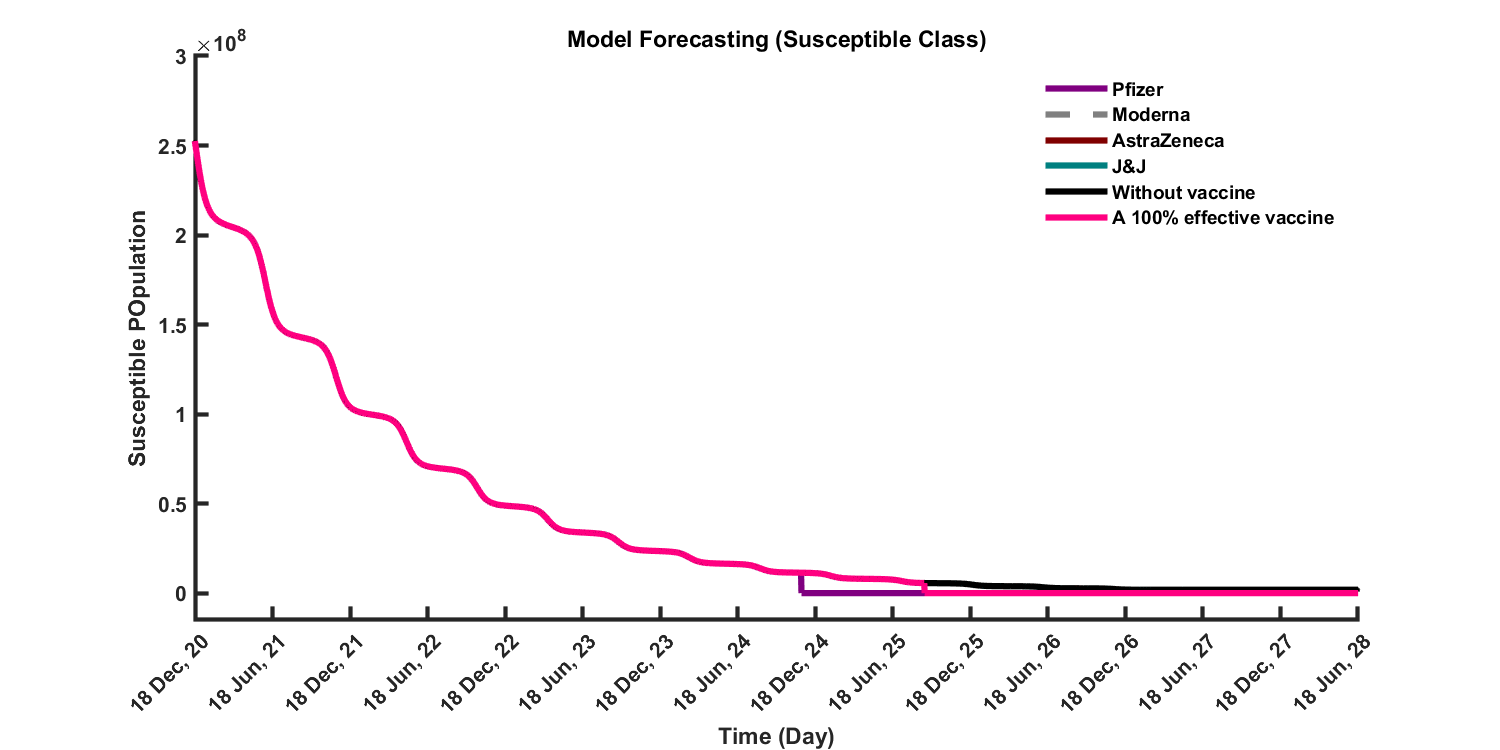}}
	\subfloat[\label{fu1}]{\includegraphics[width = 0.5 \linewidth]{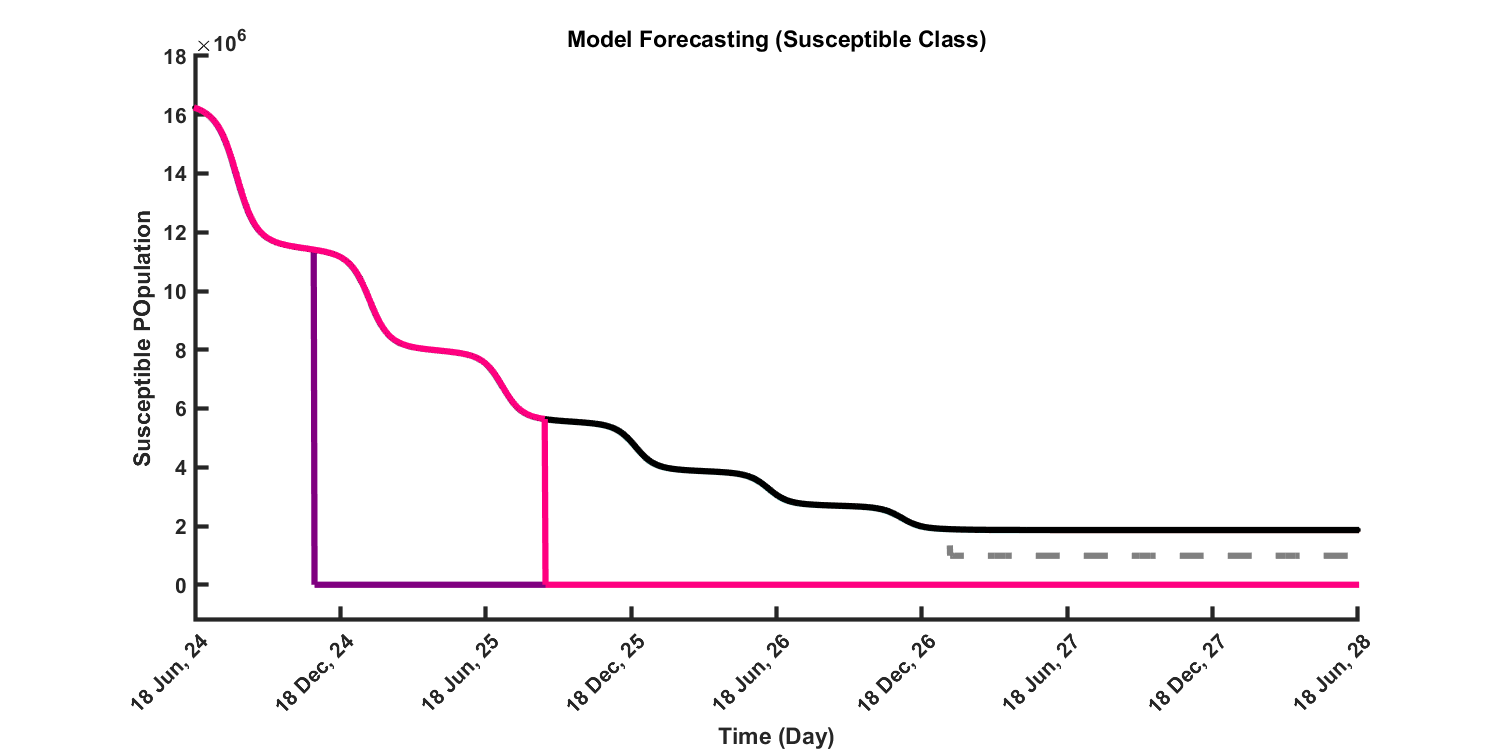}} \\
\end{figure}

\begin{figure} [H] \ContinuedFloat
	\centering
	\subfloat[\label{fu2}]{\includegraphics[width = 0.55 \linewidth]{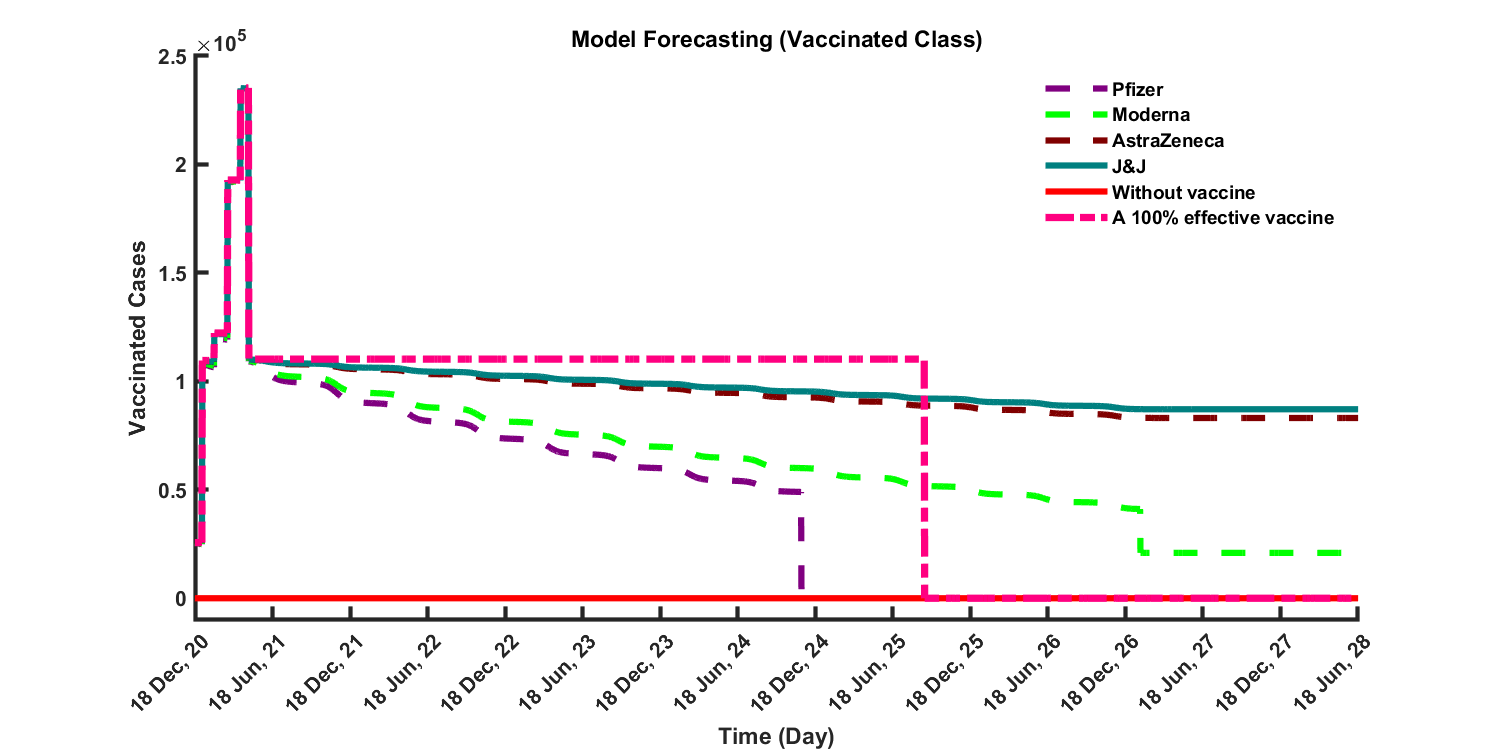}}
	\subfloat[\label{fu3}]{\includegraphics[width = 0.5 \linewidth]{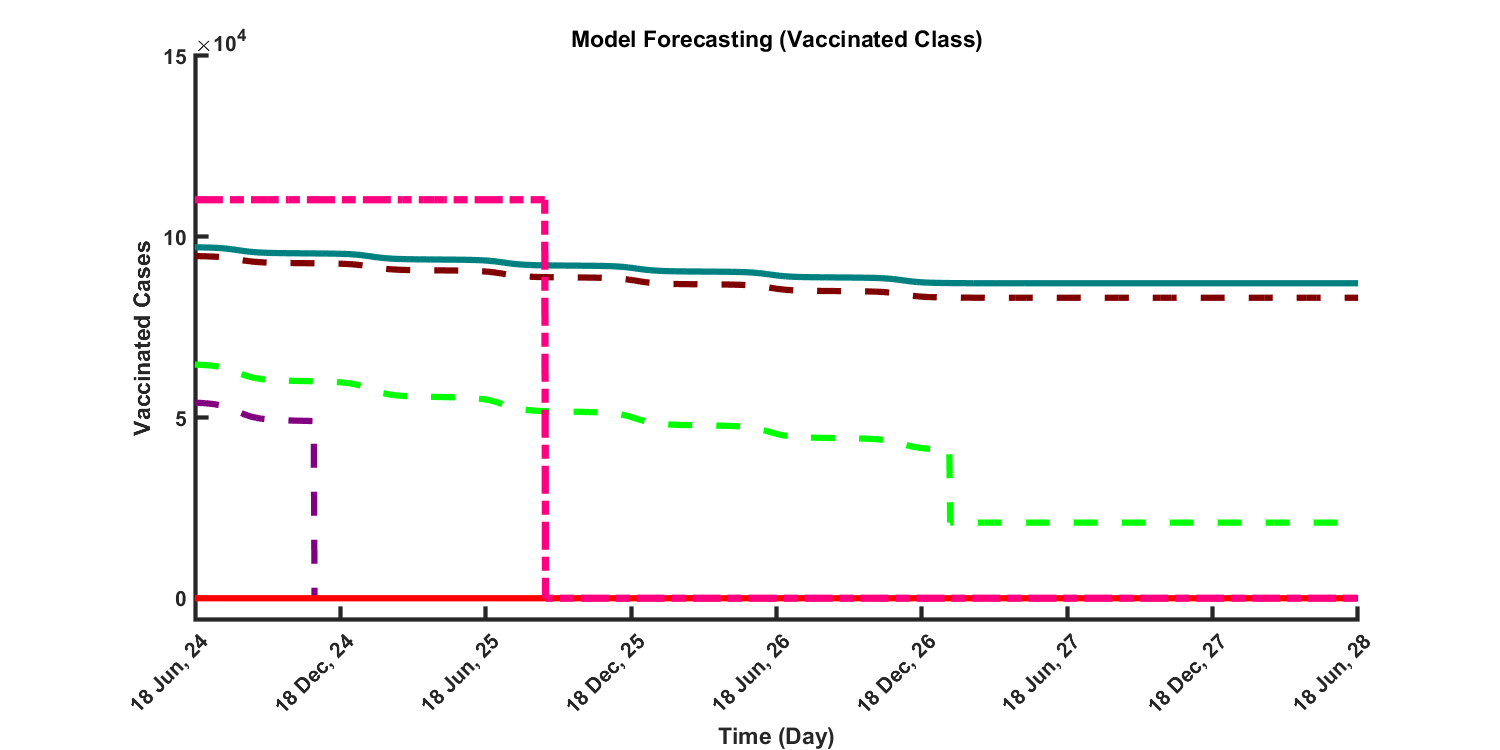}} \\
\end{figure}

\begin{figure} [H] \ContinuedFloat
	\centering
	\subfloat[\label{fu4}]{\includegraphics[width = 0.55 \linewidth]{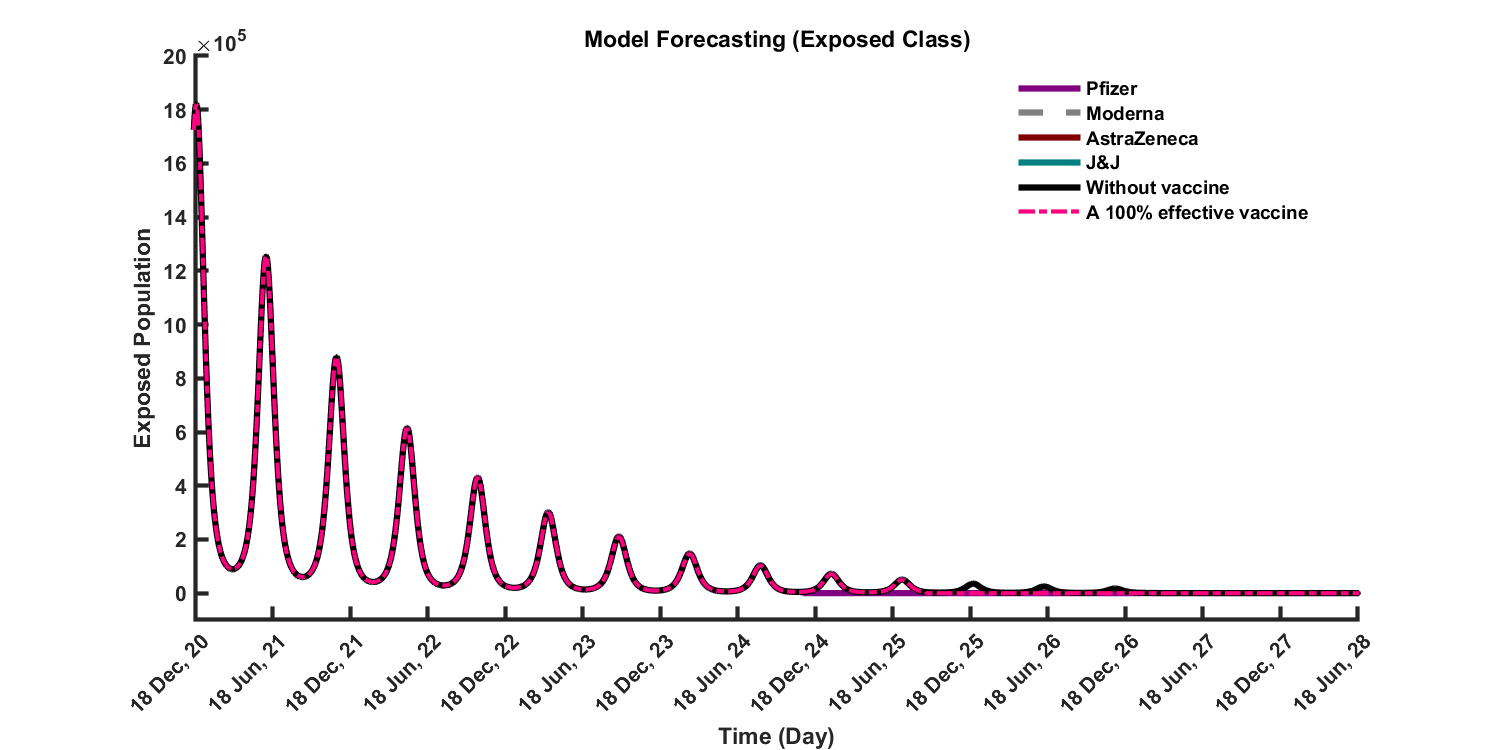}}
	\subfloat[\label{fu5}]{\includegraphics[width = 0.5 \linewidth]{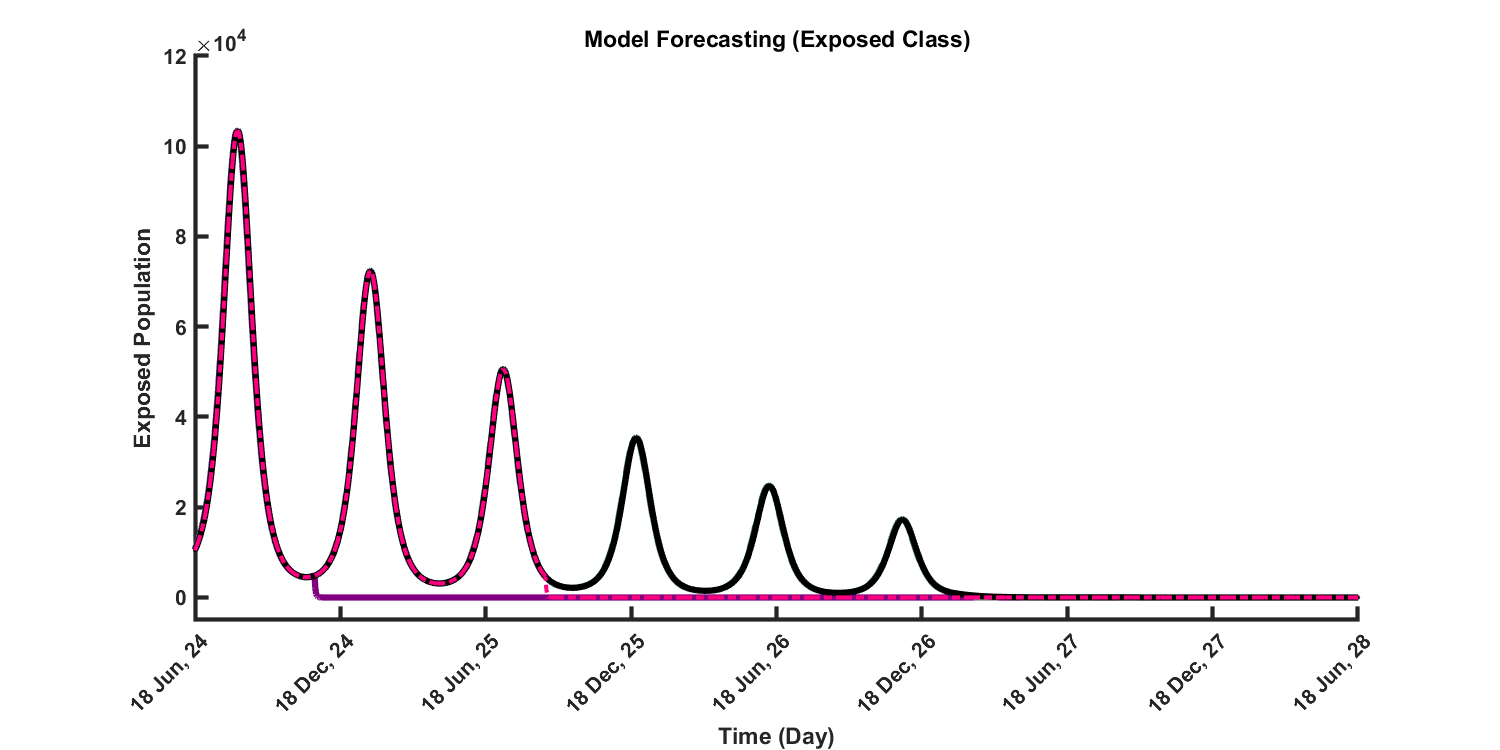}} \\
\end{figure}

\begin{figure} [H] \ContinuedFloat
	\centering
	\subfloat[\label{fu6}]{\includegraphics[width = 0.55 \linewidth]{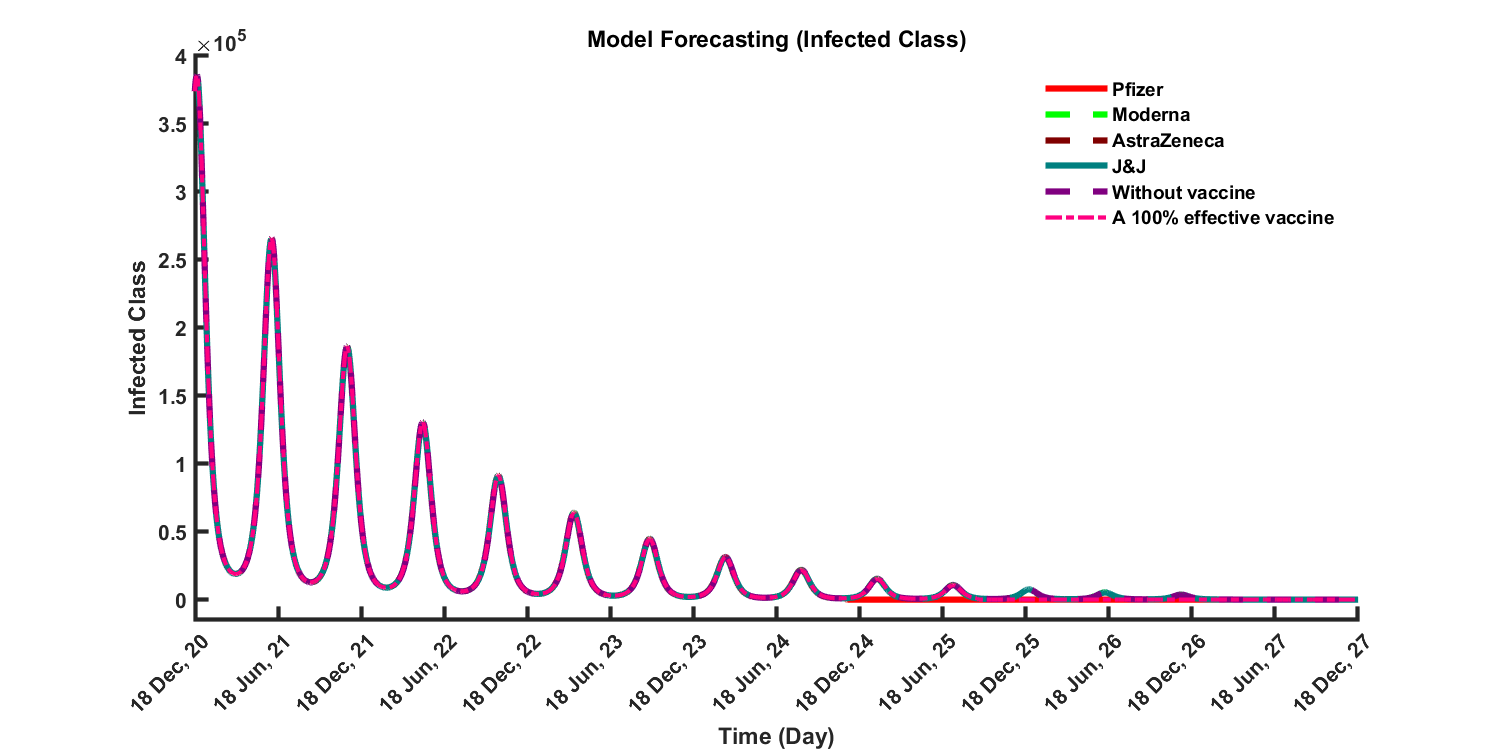}}
	\subfloat[\label{fu7}]{\includegraphics[width = 0.5 \linewidth]{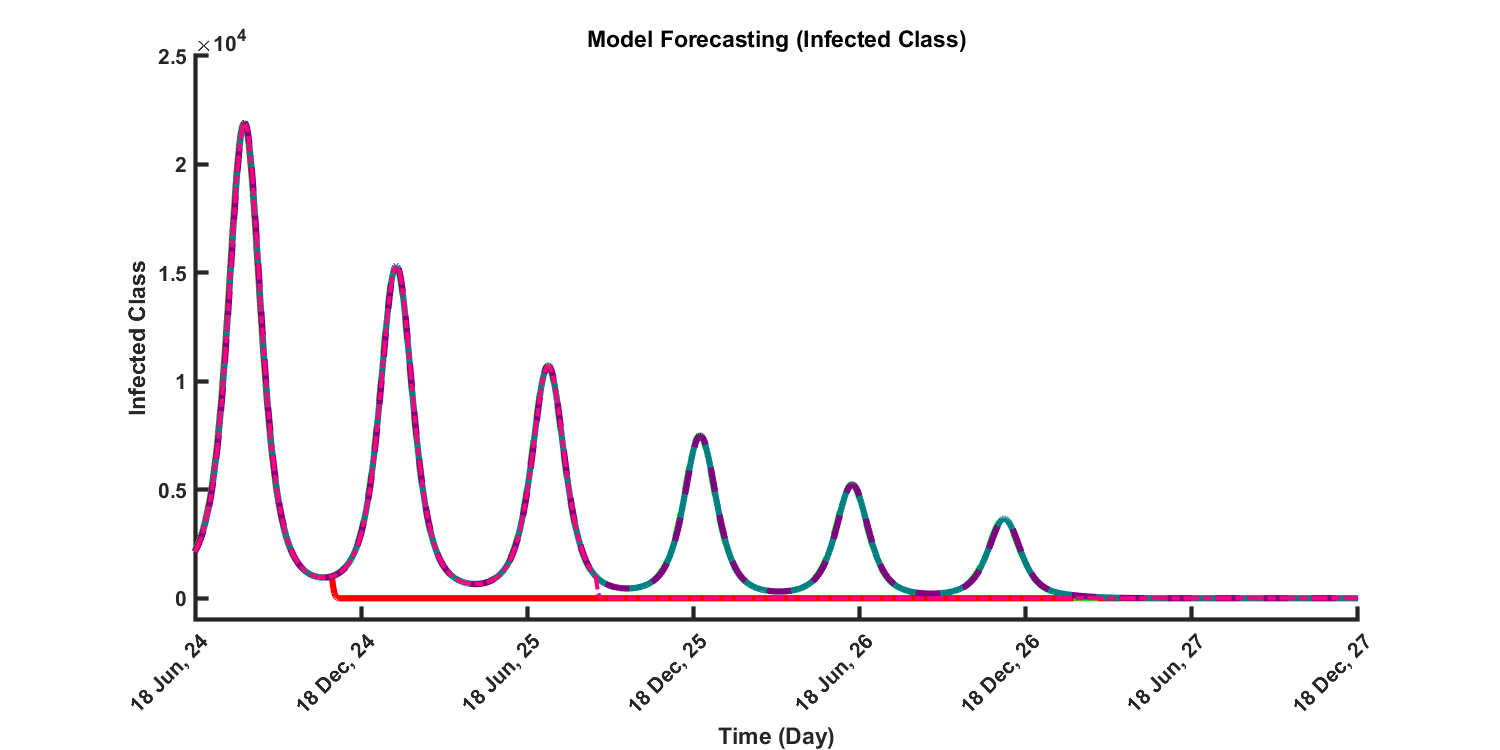}} \\
\end{figure}

\begin{figure} [H] \ContinuedFloat
	\centering
	\subfloat[\label{fu8}]{\includegraphics[width = 0.55 \linewidth]{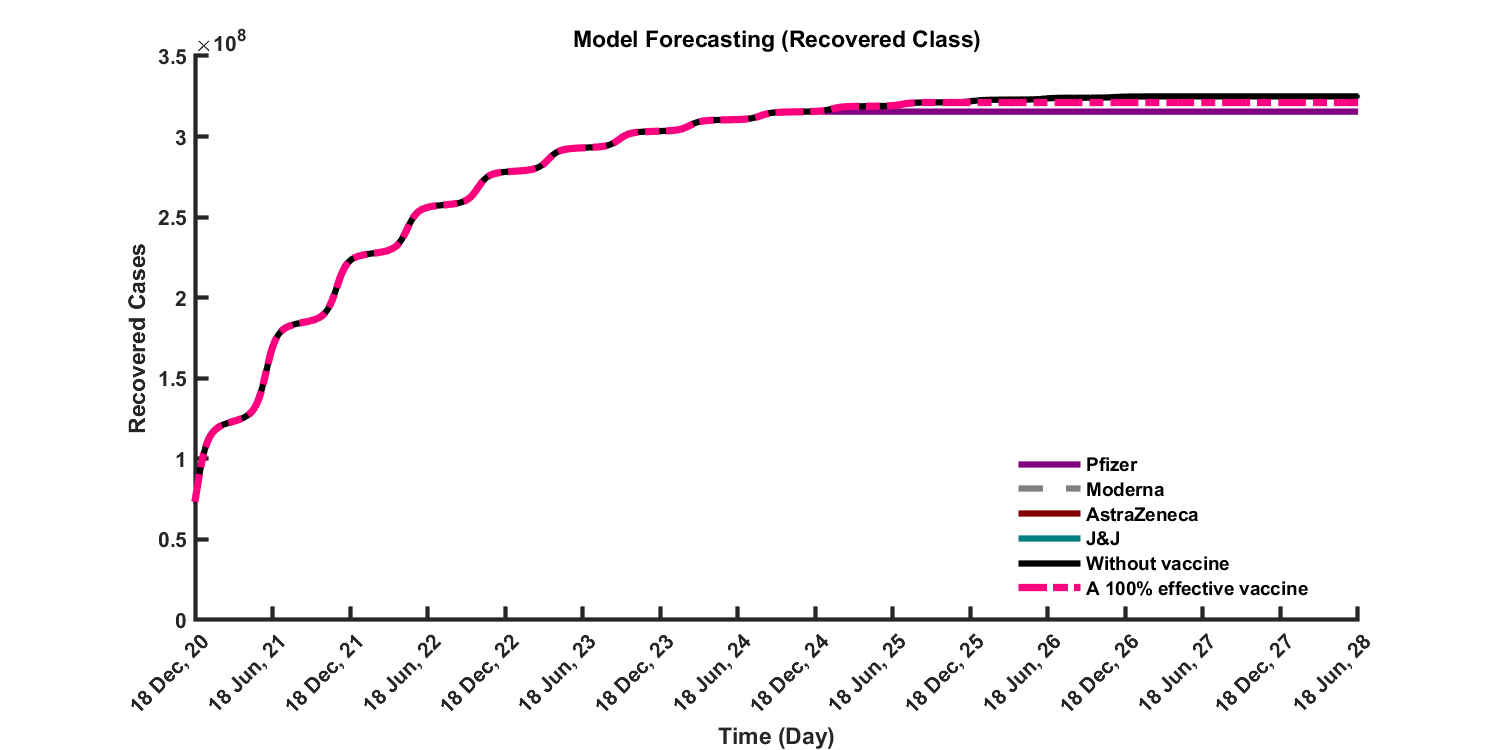}}
	\subfloat[\label{fu9}]{\includegraphics[width = 0.5 \linewidth]{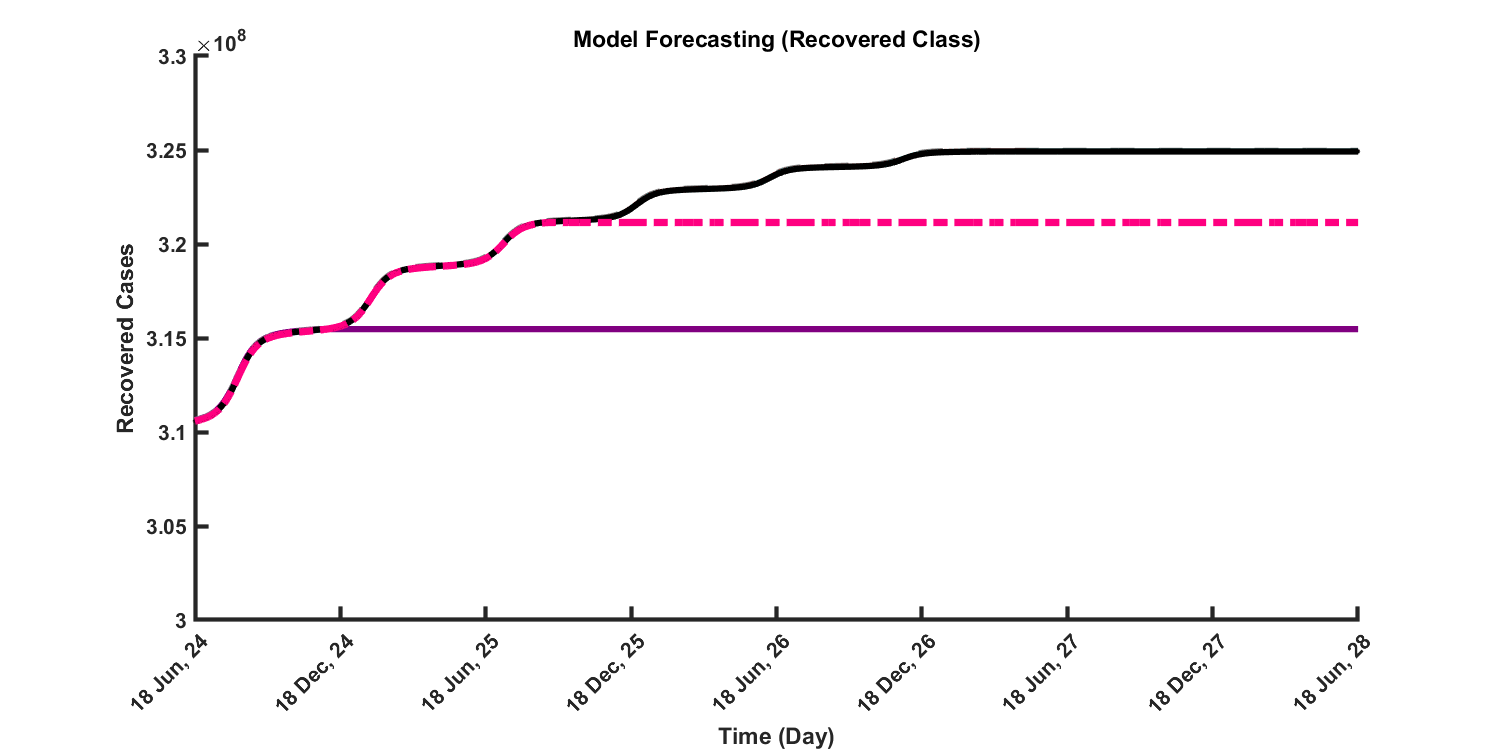}} \\
	\caption{Model forecasting against all available vaccines for the U.S. pandemic situation.}
	\label{com_fore_us}
\end{figure}

\end{document}